\documentclass[traditabstract]{aa}  

\usepackage{graphicx}

\usepackage{txfonts}
\usepackage{pdflscape}
\usepackage{aalongtable}
\usepackage{multirow}
\usepackage{natbib}

\begin{document} 

\title{The VISTA Orion mini-survey: 
       \\
      star formation in the Lynds 1630 North cloud 
  \thanks{Based on observations collected at the ESO La Silla Paranal 
Observatory under programme ID 060.A-9285(B)} }

\titlerunning{The VISTA survey in L1630\,N}

\author{ 
L. Spezzi\inst{1,2}, M. G. Petr-Gotzens\inst{1}, J. M. Alcal\'a\inst{3}, 
J. K.~J{\o}rgensen\inst{4}, T. Stanke\inst{1}, M. Lombardi\inst{5,6}, J. F. Alves\inst{7}}
\authorrunning{Spezzi,  Petr-Gotzens, Alcal\'a, et al.}

\institute{
 European Southern Observatory, Karl-Schwarzschild-Stra{\ss}e 2, 85748 Garching bei M\"unchen, Germany \email{loredana.spezzi@gmail.com}
          \and 
 {European Organisation for the Exploitation of Meteorological Satellites (EUMETSAT), Eumetsat Allee 1, 64295 Darmstadt, Germany}
    \and      
 {INAF - Osservatorio Astronomico di Capodimonte, via Moiariello, 16, 80131 Napoli, Italy } 
          \and
 {Niels Bohr Institute, University of Copenhagen, Juliane Maries Vej 30, DK-2100 Copenhagen {\O}, Denmark } 
        \and 
{University of Milan, Department of Physics, via Celoria 16, 20133 Milan, Italy}
       \and
{Harvard-Smithsonian Center for Astrophysics, Mail Stop 72, 60 Garden Street, Cambridge, MA 02138, USA}
  \and
{Institute for Astronomy, University of Vienna, T\"urkenschanzstr.\ 17, A-1180 Vienna, Austria}
}

  \offprints{mpetr@eso.org}
   \date{Received ...; accepted ...}

\abstract
{The Orion cloud complex presents a variety of star formation mechanisms and properties and it is still one of the most
intriguing targets for star formation studies.
We present VISTA/VIRCAM near-infrared observations of the L1630N star forming region, including the stellar clusters 
NGC\,2068 and NGC\,2071, in the Orion molecular cloud B and
discuss them in combination with Spitzer data. We select 186 young stellar object (YSO) candidates in the region on the 
basis of multi-colour criteria, confirm the YSO nature of the majority of them using published spectroscopy from the 
literature, and use this sample to investigate the overall star formation properties in L1630N.
The K-band luminosity function of L1630N is remarkably similar to that of the Trapezium cluster, i.e., it presents a
broad peak in the range 0.3-0.7 M$_\odot$ and a fraction of sub-stellar objects of $\sim$20\%. The fraction of YSOs still 
surrounded by disk/envelopes is very high ($\sim$85\%) compared to other star forming regions of 
similar age (1-2\,Myr), but includes some uncertain corrections for diskless YSOs. Yet, a possibly high disk fraction
together with the fact that 1/3 of the cloud mass has a gas surface density above 
the threshold for star formation ($\sim$129 M$_\odot$ pc$^{-2}$), points towards a still on-going star formation activity 
in L1630N. The star formation efficiency (SFE), star formation rate (SFR) and density of star formation of L1630N 
are within the ranges estimated for galactic star forming regions by the Spitzer "core to disk" and "Gould's Belt" 
surveys. However, the SFE and SFR are lower than the average value measured in the Orion A cloud and, 
in particular, lower than that in the southern regions of L1630. This might suggest different star formation 
mechanisms within the L1630 cloud complex. 
}
  \maketitle
   
\section{Introduction}

Observations of young stellar populations in nearby star forming regions are important tools to understand
the interplay between the outcome of the star formation process and the original environment from
which the stellar ensembles emerged. While details of the star formation process and its physics are 
often tested with targeted investigations on small spatial scales, global properties are best assessed 
with wide-field imaging surveys in the infrared thereby accomplishing large-scale studies in a homogenous 
way.

The {\it Spitzer} \emph{c2d} \citep{Eva09} and {\it Spitzer} Gould Belt (GB)\footnote{http://www.cfa.harvard.edu/gouldbelt} 
Legacy surveys \citep[e.g.,][]{Spe11,Hat12,Dun13} effectively traced the population 
of young stellar objects (YSOs) in several nearby star forming regions. These studies have shown that current star-formation 
efficiencies are in the range from 3\% to 6\%, and that star formation is highly concentrated to regions of high 
extinction with the youngest objects being strongly associated with dense cores. The great majority (90\%) of the young 
stars lie within loose clusters with at least 35 members and a stellar density of 1~M$_\odot \, $pc$^{-3}$ 
\citep[][and references therein]{Eva09}. 
The c2d and GB surveys have also shown that the star-formation surface density in galactic star forming regions 
is more than an order of magnitude larger than predicted from extragalactic star formation rate $-$ gas relationships, 
e.g. the Kennicutt-Schmidt law \citep{Eva09, Hei10}. 

Among the most-studied nearby active star formation sites are the Orion~A and Orion~B molecular clouds.
The clouds have similar masses of a few $10^4$\,M$_{\sun}$ and appear
physically connected, indicating that they stem from the same overall giant molecular cloud 
complex. However, star formation differs quite significantly between the clouds.  
In Orion~B almost all stars ($\sim$90\%) formed in stellar clusters (Lada et al.\ 1991), which 
concentrate at two major sites, one in the southern part of the Orion~B cloud
where the clusters NGC2024/23 are located and one in the northern part of Orion B (also named L1630N)
with the clusters NGC2068/71. Orion~A, on the other hand, shows a substantial population of distributed
star formation with $\sim70$\% of the stars forming in isolation (Strom et al.\ 1993, Fang et al.\ 2009),
with the exception of the Orion Nebula Cluster which lies at the northernmost end of Orion~A.
Orion~B contains several early B-type stars and at least one O star, while Orion A (excluding the ONC) possibly has no
stars earlier than B4 and is apparently deficient in early type massive stars when compared to its known 
numbers of low-mass stars \citep{Hsu12}. 
Furthermore, large-scale molecular gas maps indicate clear substructure on scales $<$2\,pc in Orion~A, 
whereas Orion~B displays very little substructure, although highly filamentary molecular gas seems
associated with the star forming regions in the northern part of Orion~B, i.e.\ in L1630\,N \citep[][and references therein]{Gib08}.

In this paper we present multi-band wide-field near-infrared observations obtained with VISTA/VIRCAM, 
combined with mid-infrared data from Spitzer, covering  approximately 1.6 square degrees in the 
northern part of the Orion molecular cloud B, i.e.\ in L1630\,N 
 
The L1630\,N region contains the prominent bright optical reflection nebulosities NGC\,2068 and NGC\,2071
which, observed at near-infrared wavelengths, reveal their full nature as young stellar clusters. \citet{Fla08} 
determined the clusters' age as 1-2\,Myr, depending on the models used, and confirmed 67 stellar members through 
optical spectroscopy. However, the use of optical spectroscopy, combined with 2MASS photometry, limited their study to 
the least embedded and slightly higher mass objects. \citet{Fang09} performed also optical spectroscopy of 132 stars in the 
region. This latter sample includes all the stars previously caracterized by \citet{Fla08} and several additional
objects classified as Pre-Main Sequence (PMS) stars by \citet{Fang09}. These authors find a much higher
disk frequency in L1630\,N in comparison with L1641 (Orion A) and with other star forming regions of similar 
age like Chamaeleon~I and IC~348, but they also caution that the results are upper limits as their sample
is biased against non-disk bearing young stars. A recent study by  \citet{Hsu12}  employed a
photometric and spectroscopic survey to enlarge the population of confirmed members in L1641 and
find the disk frequency similarly high as for L1630. Clearly, the physical characterisation of young star 
properties gets better defined
as our census of the young star populations becomes complete.

The combination of wide field coverage and excellent sensitivity of our survey enables us to uncover 
a large young stellar object (YSO) population in L1630\,N, which increases the number of previously
known YSOs by a factor 1.5, and thereby allows us to analyse the global star formation properties 
in this region. 

After the description of the VISTA-Orion catalog and the extraction of the data for this work in Sect.\ 2, 
we present in Sect.\ 3 the selection procedure of the YSO candidates in L1630\,N. Then, in Sect.\ 4 and 5 we 
investigate the K-band luminosity function, initial mass function and proto-planetary disk fraction for 
the identified sample of YSO candidate members in L1630\,N, as well as compare our results to other nearby 
young stellar clusters and associations.  We study the spatial distribution and clustering properties of YSOs and independently 
confirm the known stellar clusters NGC\,2068 and NGC\,2071, but also identify a new stellar group around 
the Herbig-Haro objects HH24-26 (Sect.\ 6). In combination with an extinction map, derived from the same 
VISTA data, we also present the results on the global properties of star formation in the region and in 
the identified sub-structures (Sect.\ 7). Our conclusions are presented in Sect.\ 8.

\section{Observations and data reduction}

\subsection{VISTA data reduction and catalog extraction \label{vista_red}}

\begin{figure*}
\centering
\includegraphics[width=15cm]{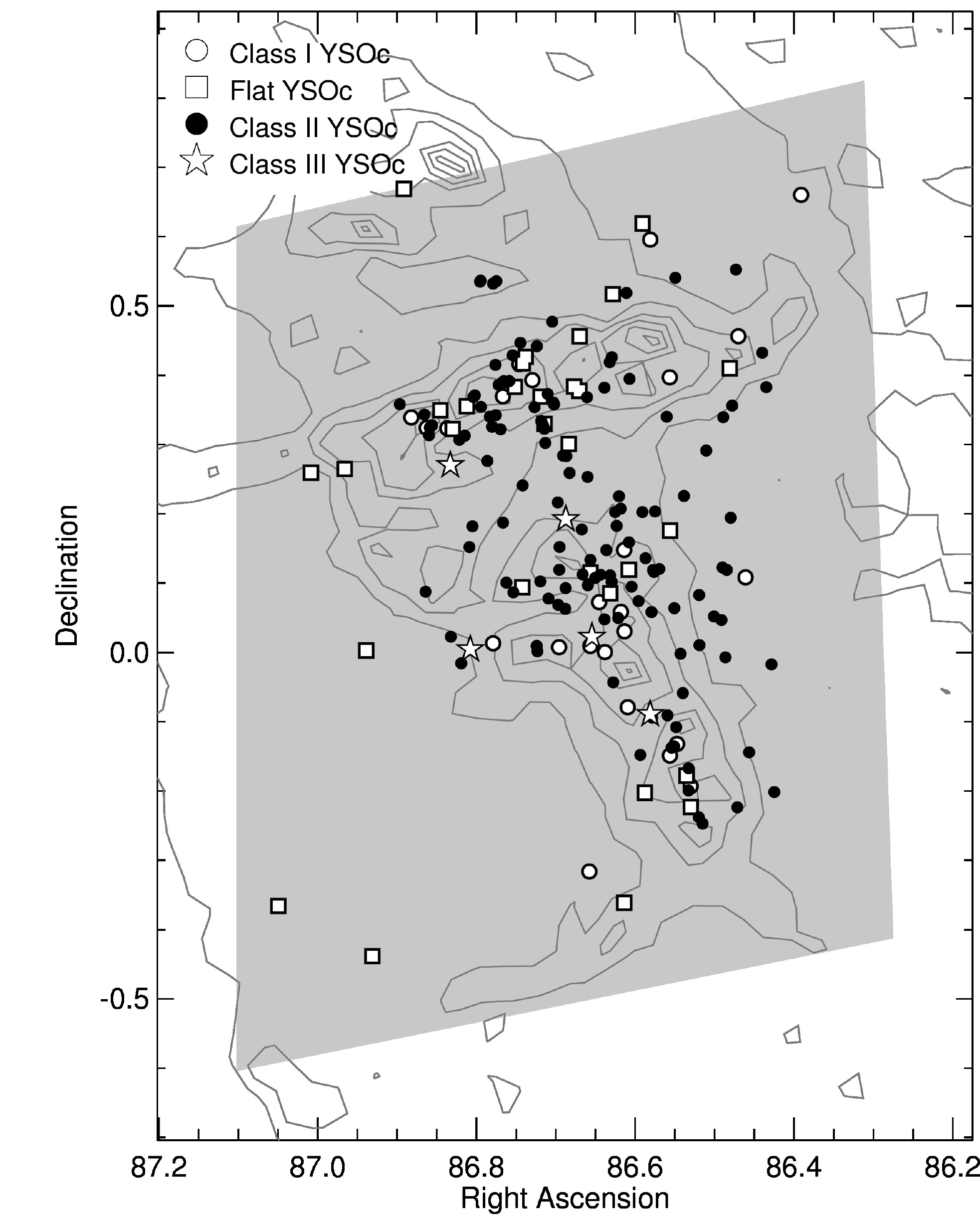}
\caption{Spatial distribution of the YSO candidates as a function of Lada classes over-plotted on the contours 
from the VISTA extinction map (solid lines). The contour levels of extinction (A$_K$) are from 0.01 to 2 mag, 
in steps of 0.2 mag. The shaded area displays the regions observed with VISTA (tile no.~12).}
\label{spa_distr}
\end{figure*}

\begin{figure*}
\centering
\includegraphics[width=13cm]{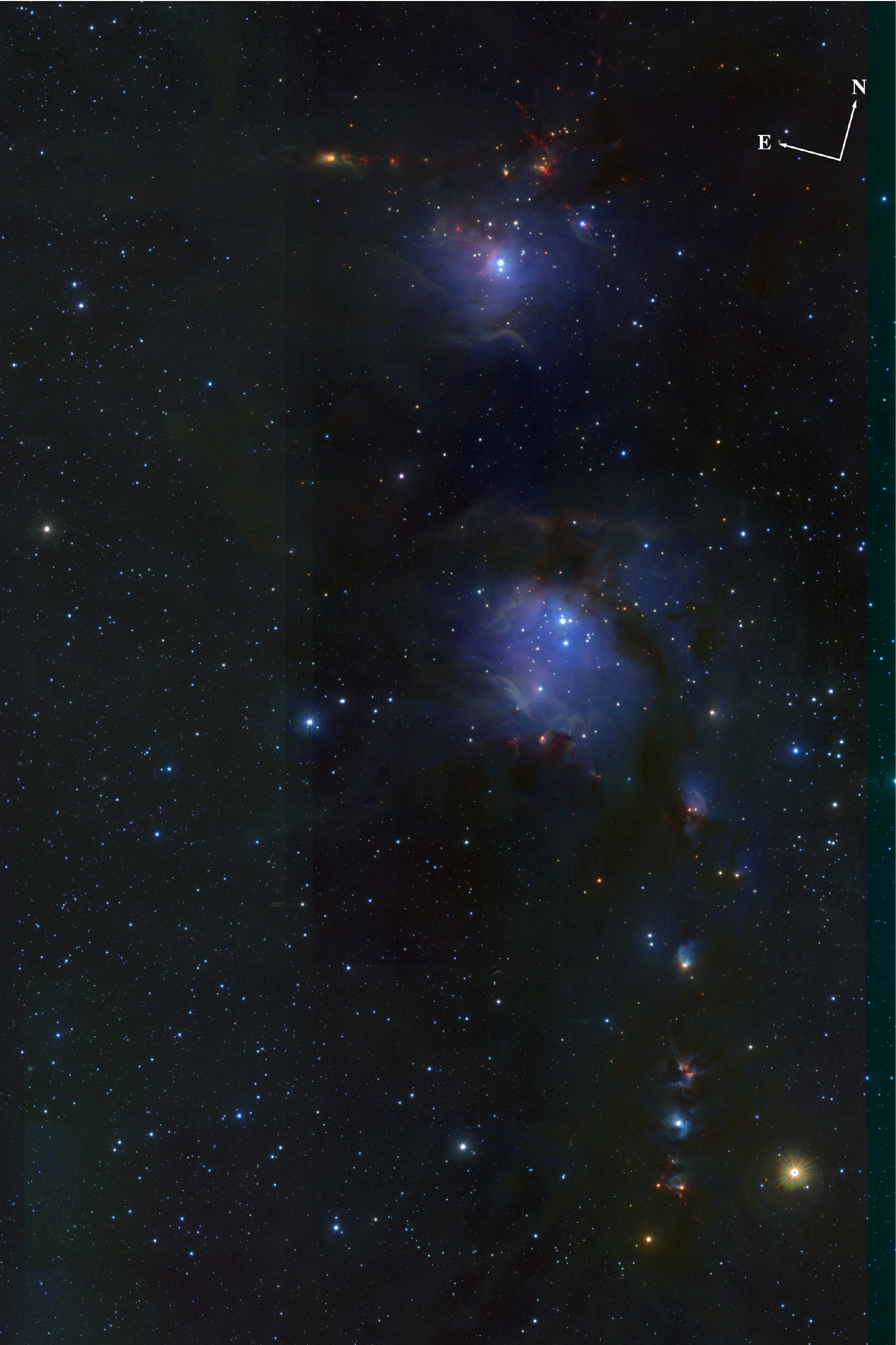}
\caption{VISTA three colour (ZJKs) mosaic image of L1630\,N on a logarithmic display. The image covers the grey shaded area of Figure~\ref{spa_distr}.}
\label{imageL1630N}
\end{figure*}

The Visible and Infrared Survey Telescope for Astronomy (VISTA) located at ESO Paranal Observatory 
is a 4m class telescope equipped with a near-IR camera (VIRCAM) containing 16 detectors, for a total 
FoV of 1\degr$ \times$1.5\degr and a pixel scale of 0.339\arcsec/pix, and available broad and narrow 
band filters in the wavelength range 0.9-2.2$\mu$m \citep{Eme06,Dal06}. Data for L1630\,N  were taken 
during the VISTA Science Verification (SV) as part of the program 
``VISTA SV Galactic Mini-survey in Orion'' \citep[PI: M. Petr-Gotzens;][]{Pet11}. 
This survey consists of $ZYJHK_S$ images obtained during 14 nights between 16 October and 2 November 2009. 
The survey area is a mosaic of 20 VISTA fields with each field containing 6 pointings that are mosaicked 
together to form a so-called filled tile. The total survey covers $\sim$30 square degrees around the 
Orion Belt stars. L1630\,N is located in the VISTA Orion survey tile no.~12 roughly centered at 
R.A.=$05^h46^m41^s$, Dec.=$+00^d09\arcmin 00\arcsec$ (Figure~\ref{spa_distr}), and contains the
young stellar clusters NGC\,2068 and NGC\,2071 which clearly stand out on the VISTA near-infrared image
(Figure~\ref{imageL1630N}).
Further details on the 
observing strategy, the exposure times per filter and particular observing patterns chosen for the 
VISTA  Mini-survey in Orion were described in \citet{Arn10} and \citet{Pet11}.

\begin{table}
\caption{Filter central wavelength, saturation limit, limiting magnitude at the 5$\sigma$ level and completeness limit for the 
        photometry of sources in tile no.~12 (L1630\,N). }           
\label{tab_obs}      
\centering                       
\begin{tabular}{ccccc}      
\hline\hline               
Filter & $\lambda_C$ & Saturation &  Mag$_{5\sigma}$ & Completeness  \\   
         &  ($\mu$m) &  limit     &                  &      limit \\   
\hline   
$Z$      &     	0.877   &  13.5 & 22.5 & 22.3 \\                     
$Y$      &      1.020   &  12.0 & 21.1 & 21.5 \\  
$J$      &      1.252   &  11.0 & 20.3 & 20.5 \\   
$H$      &      1.645   &  11.0 & 19.3 & 19.5 \\   
$K_S$    &      2.147   &  10.0 & 18.5 & 18.5 \\   
\hline                                   
\end{tabular}
\end{table}

\begin{figure}
\centering
\includegraphics[width=9cm]{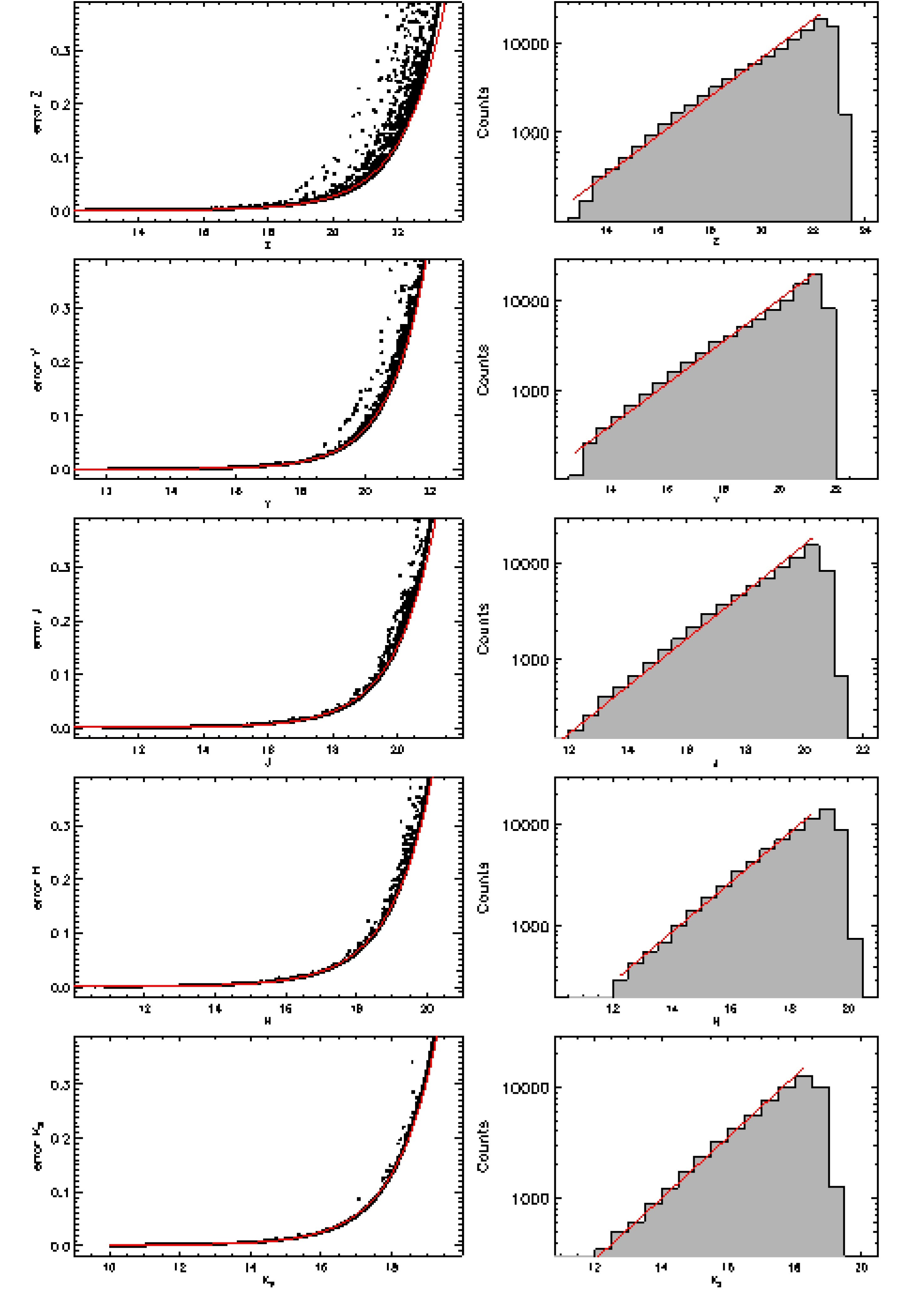}
\caption{{\bf Left panels:} Photometric errors as a function of magnitudes and relative exponential fit (continuous line) for non-saturated sources detected 
by the VISTA Orion Survey tile no.~12 in the $ZYJHK_S$ filters. 
{\bf Right panels:} Number of detection as a function of magnitude. The line shows the linear fit to the logarithmic number 
distribution of magnitudes, which is used 
to find the turning point of the distribution, indicating our completeness limit. }
\label{fig_errors}
\end{figure}

The data reduction was performed by a dedicated pipeline, developed within the VISTA Data Flow System (VDFS), 
and run by the Cambridge Astronomy Survey Unit (CASU)\footnote{http://apm49.ast.cam.ac.uk/surveys-projects/vista/vdfs}. 
The pipeline delivers science-ready stacked images and tiles, 
as well as photometrically and astrometrically calibrated source catalogues \citep{Irw04}. 
A total of $\sim$3.2 million sources were detected in the VISTA Orion Survey, and $\sim$155000 in tile no.~12 
used for this work. 
The catalog also provides, for each source in each filter, a morphological parameter (FLAG) equal to $-1$ 
for point-like sources, $1$ for extended sources, $-2$ for borderline point-like sources, and $-7,-9$ for 
problematic detections, e.g.\ sources partly saturated or whose magnitude is contaminated by bad-pixels inside 
the aperture used for the photometry extraction, or truncated because the source is very close to the mosaic 
borders. The astrometric accuracy in the source catalog is $\sim0\farcs2$ with respect to the UCAC4 
catalog \citep{Zac13}. Stellar sources typically show a FWHM of $0\farcs6-0\farcs8$.
The instrumental magnitudes are obtained through aperture photometry and 
calibrated onto the VISTA system via non-saturated 2MASS stars in the field.
Absolute photometric uncertainties are below 5\% and the 5$\sigma$ limiting magnitudes, completeness and 
saturation limits in each filter are listed in Table~\ref{tab_obs} for the specific case of tile no.~12. 
The achieved magnitude limits in $JHK_S$ are $\sim$3~mag deeper than 2MASS. 
The approximate completeness limit in each filter was derived as the point where the histogram of the 
magnitudes (Fig.~\ref{fig_errors}) diverges from the dotted line, which represents the linear fit to the 
logarithmic number of objects per magnitude bin, calculated over the intervals of good photometric 
accuracy \citep{San96,Wai92}. 
In order to estimate stellar mass limits from our saturation and completeness limits, we compare 
with the theoretical isochrones by \citet{Bar98} and 
\citet{Cha00}. Since the isochrones are provided for the Johnson-Cousins photometric system, which 
is different 
from the VISTA photometric system, we converted them to the VISTA photometric system as described
in Appendix~\ref{isocr_vista}.
We estimate that our 
survey should have detected, for a population as young as $\sim$2~Myr at a distance of about 400~pc 
(i.e., the case of L1630\,N) essentially all objects from $\sim$1~M$_\odot$ down to $\sim$5 Jupiter 
masses ($\sim$0.0045~M$_\odot$) in a region showing less than 1~mag of visual interstellar extinction.

\subsection{Spitzer data \label{spitzer_cat}}

The Orion clouds were observed by the {\it Spitzer} Space Telescope \citep{Faz04,Rie04} as part of the Guaranteed Time Observation (GTO) 
programs PID 43, 47, 50, 58, 30641, and 50070. The extraction of point source photometry from this survey in the four IRAC bands and 
the MIPS~24$\mu$m band and an overview of the basic properties of the resulting point source catalog have been presented by \citet{Meg12}.
The catalog contains 298405 sources that are detected in at least one of the IRAC bands or in the MIPS~24$\mu$m  band, the limiting 
magnitudes at the 10$\sigma$ level are roughly 16.5, 16.0, 14.0, 13.0 and 8.5 at 3.6, 4.5, 5.8, 8 and 24~$\mu$m,  respectively 
\citep[see Fig.~2 by][] {Meg12}. The catalog exhibits spatially varying completeness due to confusion with nebulosity and crowding 
of point sources in dense clusters \citep[see Fig.~3-4 by][]{Meg12}.
The Orion IRAC/MIPS maps are broken into several fields centered on the regions of strong ${}^{13}\mathrm{CO}$ emission \citep{Mie94}. 
We used a sub-set of the general catalog covering an area of 2.58~deg$^2$ around  L1630\,N.

\section{Selection of YSO candidates}

The $J-H$ vs.\ $H-K_S$ color-color (CC) diagram is traditionally used to select low-mass YSO candidates (YSOc) on the basis of near-IR data, 
because young K/M type stars exhibit a narrow range of colors in this diagram and, in particular, an $H-K_S$ excess mainly arising from 
their circumstellar disk and/or in-falling envelope  \citep{May97,Luh99,Lee05}. However, near-IR colors of YSOs may  be mimicked, to some 
extent,  by highly reddened stellar photospheres of older main-sequence dwarf stars in the field and, hence, the selected sample might 
be contaminated. On the other hand, the identification of YSOs on the basis of mid to far-IR colors alone is not trivial, because the 
YSO colors in this wavelength regime are very similar to those of many background galaxies \citep[Sect.~3.1 by][]{Eva09}. 
Thus, to select YSOs in  L1630\,N  we adopt the approach by \citet{Har07}, based on the combined use of $JHK_S$ photometry and {\it Spitzer} 
IRAC/MIPS colors. 
By using {\it Spitzer} observations of the Serpens star forming region and the SWIRE catalog of extragalactic sources  \citep{Lon03}, these authors defined 
the boundaries of the  disk-bearing YSO locus in several IRAC/MIPS color-magnitude and CC diagrams. Their criteria have proven to provide an optimal 
separation between disk-bearing YSOs (mainly class~II, class~I and Flat Spectrum objects), reddened field stars and galaxies, with the fraction of remaining contaminants
(mainly background field dMe dwarfs, a few  K/M-type giants, and Be and AGB stars) estimated to be around 30\% \citep[e.g.,][]{Spe08,Oli09,Cie10}. 
These criteria have been applied to select YSO candidates in all star forming regions observed within the frame of the {\it Spitzer} \emph{c2d} 
\citep{Eva09} and {\it Spitzer} Gould Belt\footnote{http://www.cfa.harvard.edu/gouldbelt} Legacy surveys \citep[e.g.,][]{Spe11,Hat12,Dun13}.

\begin{figure*}
\centering
\includegraphics[width=11.5cm]{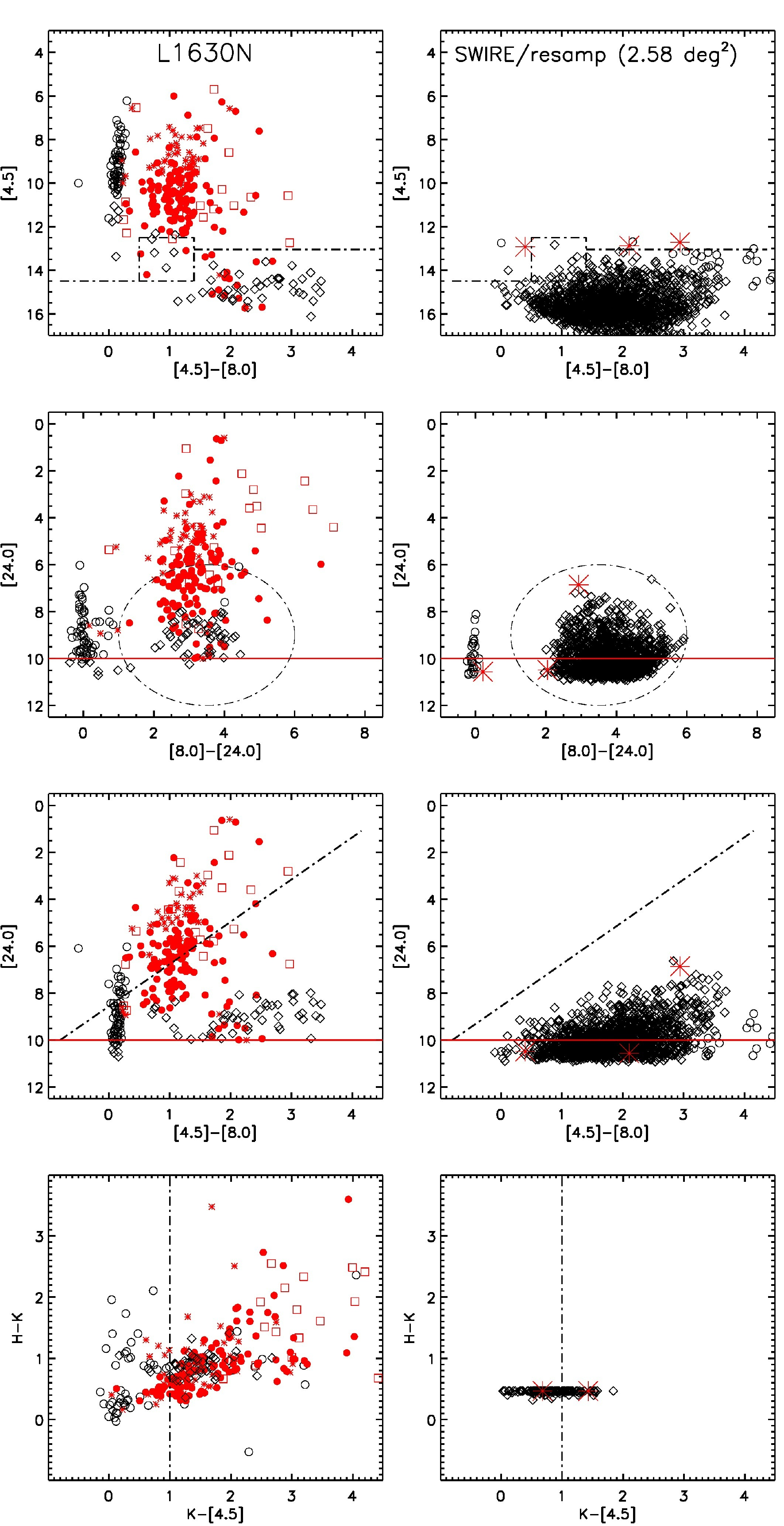}
\caption{
{\bf Left panels:} VISTA/{\it Spitzer} CM and CC diagrams for L1630\,N. The dot-dashed lines show fuzzy limits with exponential cutoffs 
that define the YSO candidate selection criterion in the each diagram, excluding contamination from galaxy (diamonds) and field stars 
presenting normal photospheric colors (circles). The continuous lines show hard limits, objects fainter than which are excluded from 
the YSO category. Point-like and extended YSO candidates are indicated by dots and squares, respectively; YSO candidates with no VISTA 
morphological classification are indicated by asterisks. {\bf Right panels:} 2MASS/{\it Spitzer} CM and CC diagrams for the SWIRE 
catalog (Sect.~\ref{vermin}). Symbols are as in the left panels. Three objects of the SWIRE catalog (marked as asterisks) are classified 
as YSOc according to our selection criteria.
}
\label{CC_sel}
\end{figure*}

We first matched our VISTA catalog for tile no.~12 with the {\it Spitzer} catalog  by \citet{Meg12} using a matching radius of 3$\arcsec$, 
larger than the astrometric accuracy of our mosaics (Sect.~\ref{vista_red}) and corresponding to twice the typical FWHM 
of sources in IRAC maps\footnote{IRAC Instrument Handbook. See http://irsa.ipac.caltech.edu/data/SPITZER/docs/irac/iracinstrumenthandbook/}. 
Then, we applied the YSO selection method by \citet{Har07} to the matched catalog, containing $\sim$58500 sources with complete $ZYJHK_S$, 
IRAC~3.6, 4.5, 5.8 and 8~$\mu$m and MIPS-24~$\mu$m photometry. A detailed review of the selection method can be found in \citet{Har07, Har07b}. 
Briefly, the selection method consists in the definition of an empirical probability function which depends on the relative position 
of a given source in several CC and CM diagrams, where diffuse boundaries have been determined to obtain an optimal separation between YSOs 
and galaxies. Figure~\ref{CC_sel} (left panels) shows the VISTA/{\it Spitzer} CC and CM diagrams used to select the YSO candidates in L1630\,N. 
Note that the method requires detection in all IRAC bands and in MIPS-24~$\mu$m with a S/N higher than 3 to classify an object as a YSO 
candidate or a background galaxy. 
Diskless YSOs, i.e.\ class~III sources, are usually rejected by the selection method.
Moreover, older field objects with no IR excess emission are rejected {\it a priori} because their IR colors are comparable with normal 
photospheric colors, e.g., $K_S-[4.5] <$-0.1,  [8]-[24]$<$0.1, [4.5]-[8]$<$0.2 \citep{Har07b}. 
In addition to the criteria by \citet{Har07}, we used the VISTA morphological parameter (FLAG; Sect.~\ref{vista_red}) to distinguish 
between point-like (FLAG=-1) and extended (FLAG=1) YSO candidates; YSO candidates truncated and/or contaminated in VISTA pass-bands 
(-9$\leq$FLAG$\leq$-2) could not be classified. 
We note that, although clearly extended sources in the VISTA mosaic are more likely to be galaxies, we can not exclude them {\it a priori}, 
because YSOs still surrounded by significant circumstellar material might appear fuzzy/extended at IR wavelengths. 
Thus, we establish the YSO or galaxy nature on the basis of the above mentioned probability function alone, which depends exclusively 
on the VISTA/{\it Spitzer} colors of the sources. 

We find 188 YSOc in  L1630\,N, shown in Figure~\ref{CC_sel} (left panels) as red dots, squares and asterisks for point-like, extended 
and morphologically unclassified sources, respectively.

\subsection{On the contamination of the YSOc sample \label{vermin}}

We  adopted a statistical method to distinguish YSOs from extragalactic contaminants and field stars, hence, 
our candidates sample could be still slightly contaminated.

The number of interloping stars can be probed by using analytic models of the Galactic stellar distribution, i.e., simulations 
of the expected properties of stars seen towards a given direction of the Galaxy over a given solid angle. We performed this 
exercise by using the Galaxy model by \citet{Rob03} and their online tool\footnote{http://model.obs-besancon.fr/}. 
In the temperature range of our candidates, foreground stars are expected to be main-sequence cool dwarfs, whereas red giants 
are expected to dominate the background population. Assuming a cluster distance of 400~pc and a typical extinction of 
A$_V$$\approx$1~mag due to the  L1630\,N cloud itself, we expect some 50 foreground dwarfs in the 1$\times$1.5~square degree 
area observed in by VISTA (tile n.~12) with apparent $K_S$ magnitude between 7 and 18.5 and spectral types of M0 to M9, 
i.e., the magnitude and spectral range corresponding to members of   L1630\,N detectable by our survey ($\lesssim$1~M$_\odot$; Sect.~\ref{vista_red}). 
Only a handful ($\sim$3) of background giants are expected to be found in the locus occupied by the cluster members because they generally 
appear much brighter than   L1630\,N members in the same effective temperature range. We thus conclude that only foreground cool 
main-sequence stars can contribute noticeably to the contamination of our candidate sample, but the contamination level is at most 25-30\%.
We note that contamination is higher ($\sim$50\%) in the substellar regime ($\sim$0.1~M$_\odot$, i.e., K$_S \approx$13~mag), and lower in the 
stellar regime ($\sim$20\%).
On the other hand, as seen in Figure~\ref{spa_distr}, the YSOc very clearly follow the cloud extinction contours which is 
not expected for a randomly distributed foreground population. Also, we find almost all of our YSOc consistent with infrared
excess sources of class~II or earlier (see Sect.\ \ref{ladaclasses}). Therefore, we conclude that the true contamination of our sample is very low.

 
In order to have a statistical estimate of possible remaining extragalactic contaminants, 
we used the {\it Spitzer} Wide-area Infrared Extragalactic \citep[SWIRE][]{Sur04} catalog coming from the observations  of the 
ELAIS~N1 field (Rowan-Robinson et al. 2004). The SWIRE catalog was trimmed and resampled as accurately as possible to match 
the spatial extent (2.58~deg$^2$) and sensitivity limits (Sect.~\ref{spitzer_cat}) of the {\it Spitzer} observations in L1630\,N.
Moreover, the photometry of sources in the SWIRE catalog was edited in order to simulate the interstellar extinction in the 
direction of L1630\,N, as expected on the basis of the VISTA extinction map (Sect.~\ref{sec_ext}). $JHK_S$ for the SWIRE sources 
are recovered from the 2MASS catalog \citep{Skr06}.
For further details on how the trimmed resampled SWIRE comparison catalog was created, we refer the reader to \citet{Eva08} 
and \citet{Har07}.
The selection criteria applied to the comparison resampled SWIRE data lead to the conclusion that only 3 (i.e., less than 2\%) 
of the selected YSO candidates in   L1630\,N     may be background galaxies (Fig.~\ref{CC_sel}). 
This result is similar to what \citet{Har07} and \citet{Alc08} found for the Serpens and Cha~II molecular clouds, respectively.
Indeed, all the 188 candidates have been visually inspected in the VISTA images and we find that only two of them are clearly 
galaxies; these two candidates have been neglected in the subsequent analysis. All the remaining 186 candidates appear 
point-like or almost point-like in all our images and their photometry is not contaminated by nearby saturated stars, or any 
other artifact that might affect our selection criterion. 
A few of them ($\sim$3\%) present a close companion in the VISTA images not resolved in the {\it Spitzer} images and, hence, 
their {\it Spitzer} fluxes might be contaminated. We can not discharge {\it a priori} these candidates, because one or both 
objects in the system might still be young and, hence, responsible for the observed IR excess emission.

This leads us to a remaining caveat in our selection method. Our YSO candidates might be members of binary/multiple 
systems too close to be resolved with VISTA/{\it Spitzer} and, hence, affecting the measured photometry. As seen in 
Sect.~\ref{vista_red}, our candidates are expected to have masses  $\lesssim$1~M$_\odot$. 
The multiplicity fraction for stars in this mass regime is estimated to be between 20\% and 40\%, depending on the 
actual mass of the primary star and the separation range \citep{Duq91,Mas98,Bas06,Lad06}. However, higher resolution 
imaging or spectroscopy would be needed to assess the actual multiplicity fraction in L1630\,N. 

\subsection{Comparison with previous surveys \label{preioussurveys} } 
 
Before our study, a census of the young stellar population in L1630\,N was presented by \citet{Fla08} and by \citet{Fang09}.
\citet{Fla08} identified 69 cluster members with a rather spread spatial distribution, i.e., not confined to regions of dense 
gas and dust. For 67 of these members, they derived accurate spectral type and luminosity and estimated a median age of 2 Myr, 
and a large fraction of stars with infrared excess actively accreting (79\%). Using a mix of criteria (presence of H$\alpha$ 
emission, \ion{Li}{I} absorption or IR excess) \citet{Fang09} selected and analized 132 PMS stars in the general direction of 
the clusters. This list includes practically all the PMS stars studied by \citet{Fla08}. For 111 stars \citet{Fang09} provide 
a classificaton in terms of their IR-excess as "thick disk, transitional disk, thin disk and no disk". Most of the 21 objects 
missing IR classification lack any information on \ion{Li}{I} absorption and were selected as PMS stars by \citet{Fang09} only 
because H$\alpha$ is detected in emission, although rather weak considering their spectral types \citep[][]{WhiteBasri03}.

Our YSOc sample consists of 186 objects.
In Table~\ref{tab_yso} we report their coordinates and VISTA + {\it Spitzer} photometry.
The VISTA/{\it Spitzer} selection criteria recovered 50 out of the 69 (i.e. $\sim$75\%) PMS stars listed by \citet{Fla08}, 
specifically 10 weak T~Tauri stars (WTTs) and 40 classical T~Tauri stars (CTTs). Likewise, our criteria recover 82 of the 
132 objects in \citet{Fang09} (i.e., $\sim$62\% of the whole sample, but 74\% of the sample with IR classification).
Most of the recovered objects by the VISTA/{\it Spitzer} selection criteria in both catalogs can be classified 
as Class~II YSOs. Our survey missed about 50 of the previously known YSOs in the area,
i.e. $\sim$21\% of the YSO population (50 of 186+50), the vast majority of which are
WTTs or objects without disks, i.e.\ Class~III YSOs. We mark in Table~\ref{tab_yso} 
the YSOc already identified by \citet{Fla08} and/or \citet{Fang09}.

Although the selection criteria and color cuts presented here are different from those by \citet{Meg12}
it is also interesting to compare our results with their selection, since we gathered the Spitzer 
photometry from their catalog (see Section~\ref{spitzer_cat}). In our studied area there are 257 
sources selected by \citet{Meg12} as possible YSOs, but 75 of these lack information in at least 
one IRAC band or at 24 $\mu$m. Since our selection criteria require the detection in all IRAC bands 
and at 24 $\mu$m, we can classify 182 of the \citet{Meg12} sources in the region. With our methods we 
thus recover 162 YSOs, meaning that our criteria miss 20 of the \citet{Meg12} sources. Most of them 
are classified as possible protostar candidates by \citet{Meg12} and are distributed in regions of 
high sellar density. Thus, our criteria recover about 90\% of the \citet{Meg12} sources.

\section{Luminosity Function and Characteristic Stellar Mass}

The stellar luminosity can be used as a poor but still useful first-order proxy for mass, assuming 
that most of the stars have formed more or less at the same time.  Before using the range of luminosities to provide an 
estimate of the mass range, we determined the degree of completeness of the YSO candidate sample in L1630\,N. Our selection 
criteria ultimately rely on the VISTA and {\it Spitzer} IRAC~3.6-8$\mu$m/MIPS~24$\mu$m detection of the objects and on the 
quality of this detection. Thus, in order to investigate the expected number of low-luminosity objects and infer the 
typical mass distribution of our YSOc sample, we have to take the completeness of these two datasets into account.

\subsection{1-30$\mu$m Bolometric Luminosity Function}

In order to estimate the completeness of the {\it Spitzer} observations in the   L1630\,N, 
we used the same approach as \citet{Har07}, which has been applied to all c2d/GB clouds \citep[e.g.,][]{Alc08,Mer08,Spe11}. 

\begin{figure}
\centering
\includegraphics[width=9cm]{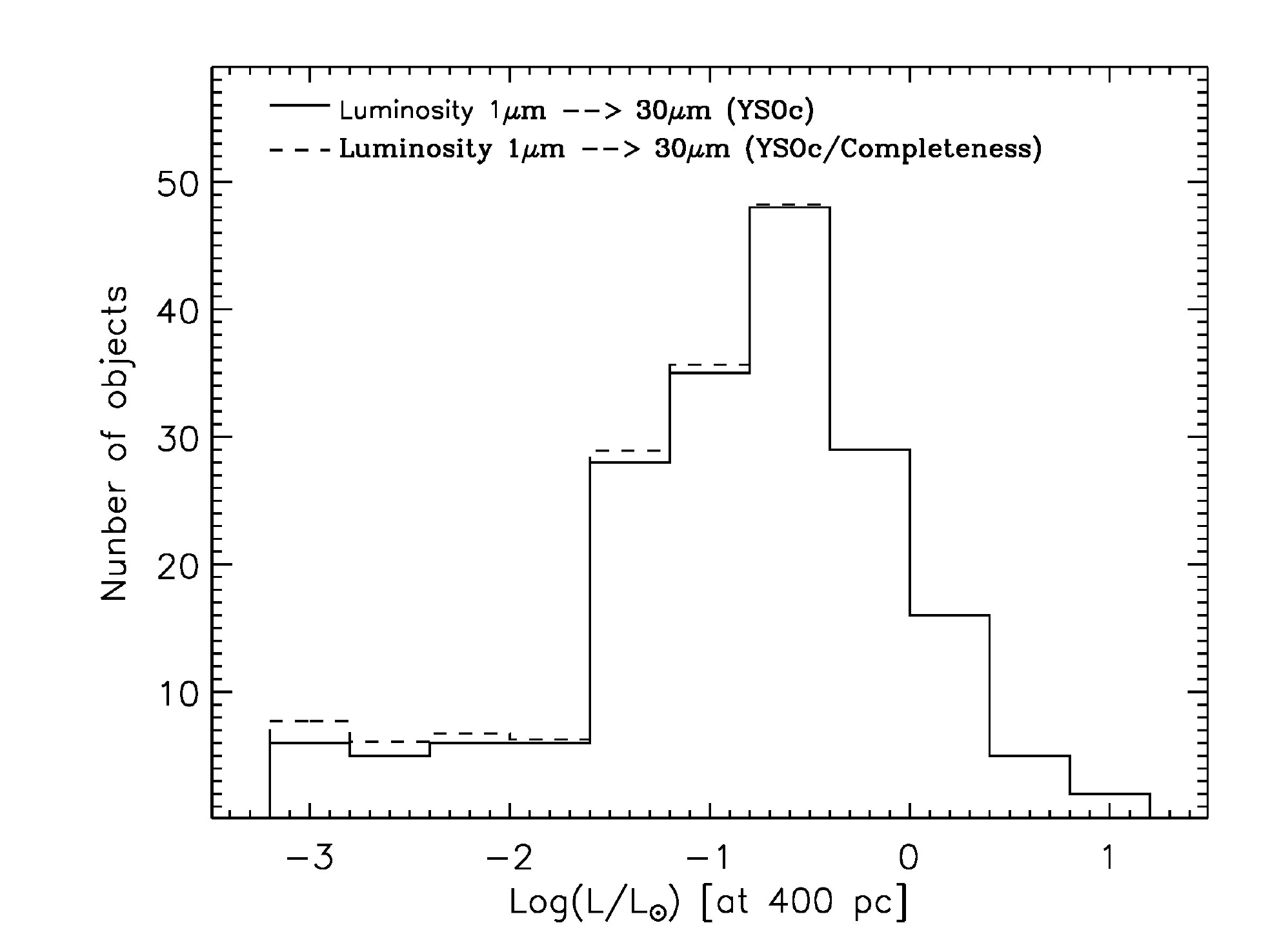}
\caption{
Luminosity distribution for the YSOc in   L1630\,N     (solid histogram). The plotted luminosities were determined as 
in \citet{Har07}, i.e., by integration of the SEDs from 1 to 30~$\mu$m. The corrected luminosity distribution, determined 
by applying completeness factors at each luminosity bin as in \citet{Har07}, is over-plotted (dashed histogram).}
\label{IR_lum}
\end{figure}

We derive first the total infrared luminosity of our YSOc by integration over their SED flux between 1 to 30~$\mu$m; the 
total flux was converted to luminosity assuming a distance of 400 pc. We then applied the \citet{Har07} completeness correction factors 
for the c2d survey to our YSO candidate samples. 
\citet{Har07}  estimated the completeness of the c2d catalogs by comparing, for each luminosity bin, the number of counts from 
a trimmed version of the deeper SWIRE catalog of extragalactic sources (assumed to represent 100\% completeness by c2d standards) 
with the number of counts for the c2d catalogs in Serpens. This completeness correction can be applied to our {\it Spitzer} catalog 
of L1630\,N  because its photometric depth (Sect.~\ref{spitzer_cat}) is similar to the one of the c2d catalogs (in both cases the 
10$\sigma$ limit is $\sim$16.5~mag for IRAC~3.6$\mu$m and $\sim$8.5~mag for MIPS~24$\mu$m; compare Fig.~23 and 26 by \citet{Eva08} 
with Fig.~2 by \citet{Meg12}). 
Figure~\ref{IR_lum} shows the 1-30$\mu$m bolometric luminosity function (BLF) for YSOc  in   L1630\,N     before (solid line) 
and after (dashed line) correction for completeness, and suggests that we are missing only a few ($<$5) of additional low-luminosity 
sources with $log(L/L_\odot)< -1.7$. 
These objects have been missed by our selection either because they are below the noise level of the {\it Spitzer} 
observations or because they are located within the galaxy loci of the CM diagrams (Fig.~\ref{CC_sel}). 
In conclusion, our YSO candidate samples in   L1630\,N     is fairly complete. The luminosity histogram suggests a completeness 
better than $\sim$95\% at luminosities down to 0.01~L$_\odot$, which correspond to a mass of 0.02 M$_\odot$ for 2~Myr old stars 
according to the PMS evolutionary tracks by \citet{Bar98} \& \citet{Cha00}. The peak of the luminosity function appears at 
0.25~L$_\odot$, which corresponds to a 0.4~M$_\odot$ star (i.e., spectral type M3 at an age of 2 Myr).
A very peculiar characteristics of the L1630\,N 1-30$\mu$m bolometric luminosity function is that it shows a significant 
number ($\sim$35\%) of low- and very low-luminosity objects ($log(L/L_\odot) \lesssim -1$). A similar tail of low-luminosity 
objects was noted in the Lupus~I, III and IV star forming regions \citep[$\sim$40\%][]{Mer08,Com08}, while it is not observed 
in other c2d/GB clouds such as Cha~II \citep[$\sim$15\%][]{Alc08}, Lupus V and VI \citep[10-20\%][]{Spe11} and Serpens \citep{Har07}.  

\subsection{K-band Luminosity Function}

To further investigate/confirm the stellar mass distribution of the  L1630\,N  YSO population, we also constructed its 
$K$-band luminosity function (KLF). We choose the KLF rather than the $J$ or $H$-band luminosity functions  in order to 
minimize the effects of extinction, to maximize our sensitivity to intrinsically red, low-luminosity members of this 
cluster, and to make detailed comparisons to the KLF of the nearby Trapezium cluster presented by \citet{Mue02}.
 We did not correct the K-band fluxes for excess emission,  but this should not affect the comparison 
with the Trapezium Cluster KLF, because \citet{Mue02} do not correct for disk excess flux neither. The majority 
of stars in the Trapezium sample are disk-sources, and the H-K color distribution of the whole
Trapezium sample is very similar to the H-K color distribution of our
YSO sample in L1630, indicating a similar disk excess nature for the samples. 

\begin{figure}
\centering
\includegraphics[width=9cm]{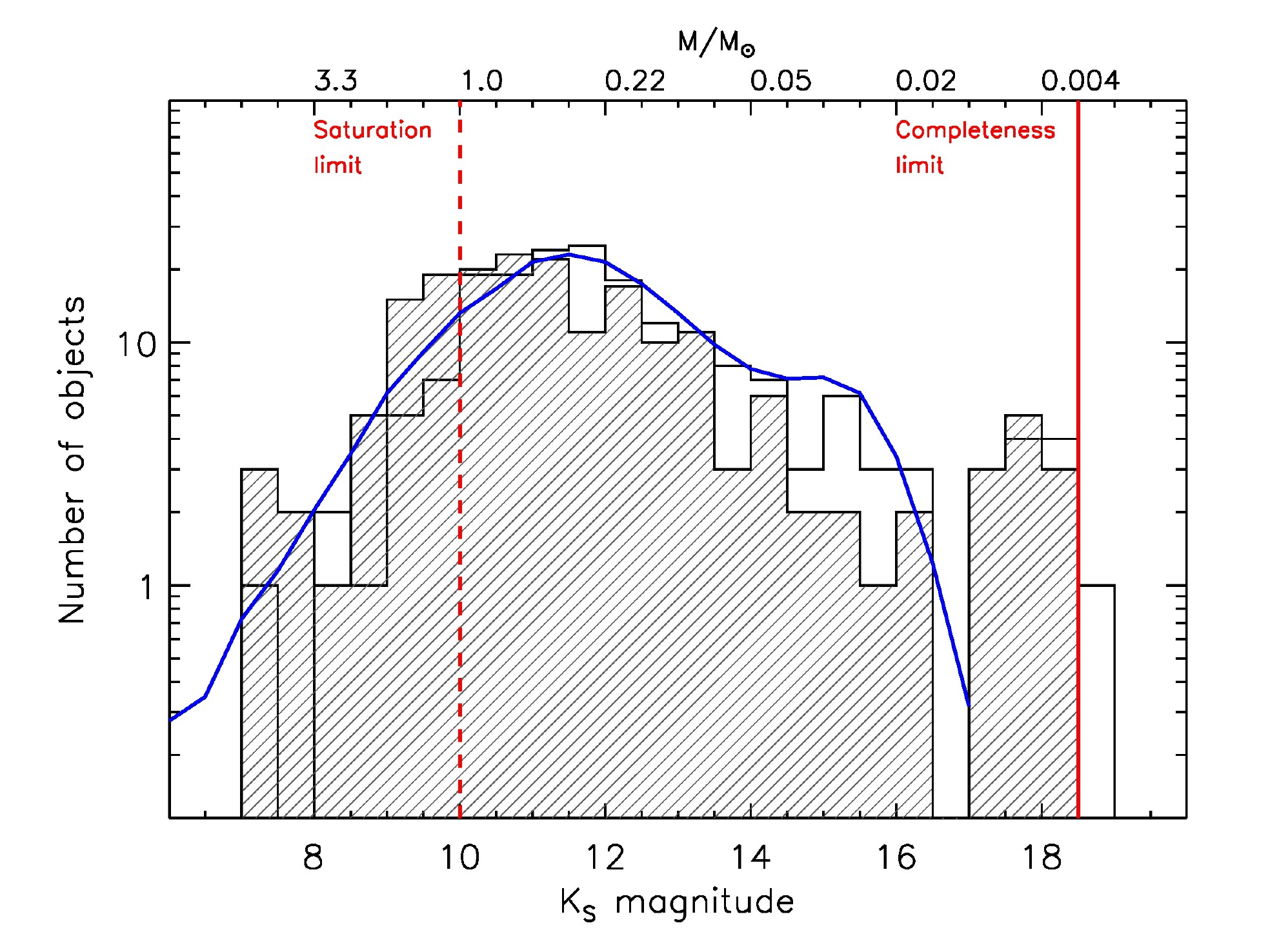}
\caption{The   L1630\,N  $K$-band luminosity function (KLF) before (empty histogram) and after correction for interstellar 
extinction (line-filled histogram). The labels in the top x-axis indicate the corresponding stellar mass according to the 
VISTA 2~Myr isochrone (Appendix~\ref{isocr_vista}). The continuous and dashed vertical lines indicate the completeness and 
the saturation limit of the VISTA $K$-band photometry, respectively. The continuous curve is the KLF of the Trapezium 
cluster \citep{Mue02}, scaled to the peak of the  L1630\,N  KLF.}
\label{KLF}
\end{figure}

In Figure~\ref{KLF} we present the observed and dereddened KLF of  L1630\,N. We use relatively wide bins (0.5~mag) that 
are much larger than the photometric errors (Fig.~\ref{fig_errors}) and adopt for each YSOc the visual extinction derived 
from the VISTA extinction map (Sect.~\ref{sec_ext}) and reported in Table~\ref{tab_yso}. 
The K-band excess could add $\sim$0.5mag, on average, to the observed
K-band magnitude \citep{May97}, i.e.\ about the same size as the KLF binsize.
However, this does not have any significant affect on our conclusions
on the KLF shape, as the excess is a property over the entire luminosity range 
(i.e.\ there is no singular effect on an individual mass regime of the KLF), and the 
steady decline in the sub-stellar regime discussed below is a robust result. We also indicate in Figure~\ref{KLF} 
the $K$-band saturation limit and limiting magnitude of our VISTA catalog (Table~\ref{tab_obs}) and overplot, for 
comparison purposes, the Trapezium KLF as derived by \citet{Mue02}, arbitrarily scaled to the peak of the L1630\,N KLF.

The KLF of L1630\,N shows a broad peak between 10.5 and 12~mag, i.e., 0.3-0.7~M$_\odot$ at the cluster distance and age 
according to the 2~Myr isochrone by \citet{Bar98} \& \citet{Cha00} converted to the VISTA photometric system 
(Appendix~\ref{isocr_vista}). Then, it steadily declines to the Hydrogen-burning limit ($\sim$0.1~M$_\odot$, i.e., K$_S \approx$13~mag).
Below this limit, we count a fraction of 28\% substellar YSOs\footnote{This fraction is computed as the number of 
substellar objects over the total number of YSOc.}. 
However, we note that the expected contamination from field stars mimicking the colors of BDs is expected 
to be $\sim$50\% (Sect.~\ref{vermin}), while it is lower in the stellar regime ($\sim$20\%) and, hence,  the actual fraction 
of sub-stellar objects in L1630\,N  could be as low as 20\%. Thus, the KLF of L1630\,N indicate a stellar mass distribution 
consistent with the 1-30$\mu$m BLF. Moreover, it appears remarkably similar to the Trapezium KLF \citep[see Fig.~11a by][]{Mue02}, 
which presents a broad peak around 0.6~M$_\odot$ and then declines into the sub-stellar regime, the fraction of sub-stellar 
objects being  $\sim$22\%. \citet{Mue02} also reported a significant secondary peak around 10-20 Jupiter masses ($\sim$0.02~M$_\odot$). 
Although we do observe a similar fraction of sub-stellar objects, the presence of this secondary peak is not obvious in the KLF 
of L1630\,N, which appears to keep its steady decline down to our completeness limit ($\sim$0.0045~M$_\odot$, i.e., K$_S \approx$18.5~mag).

The mass function shape of the Trapezium cluster in the sub-stellar regime has been long debated, because the large fraction 
of brown dwarfs (BDs) with respect to other nearby star forming regions \citep[$\sim$15\%;][]{Bri02,Lop04,Spe07,Spe08,Spe09} 
could be an affect of spatial and photometric incompleteness of the surveys conducted so far. However, more and more complete 
surveys are now available and, still, there is no universal agreement on the behavior of the mass function over the substellar 
regime. For a complete collection of BD fraction measurements in nearby star forming regions and a discussion on possible 
trends, we defer the reader to \citet{Sch12} and references therein. Note that these authors compare the star-to-BD ratio 
(R$_{star/BD}$) for various star-forming region (see their table~5). It appears more and more evident that, on one hand, 
there are clusters like IC~348, T association like Taurus, Chamaeleon, $\rho$-Ophiuchus, etc., with a low number of 
sub-stellar objects (R$_{star/BD}$ in the range 5 to 8), and on the other hand there are more massive clusters such as 
Trapezium, the Orion Nebula Cluster (ONC), NGC~1333, Upper Scorpius, etc., where this number is much higher 
(R$_{star/BD} \approx$ 2-4). L1630\,N, as other clusters in the Orion complex (Trapezium and the ONC), would belong to 
this last category, with R$_{star/BD} \approx$2.5-3.9 depending on the actual contamination level. The variation of the 
fraction of substellar objects observed from region to region possibly indicates environmental effects on their formation 
mechanism, a key point of the current star formation theory still under debate \citep[e.g.,][]{Whi07}. 

\section{Lada classes, disk fraction and disk lifetimes}
\label{ladaclasses}

In this section we revise the disk properties of the young stellar population in NGC2068/2071 on the basis of our YSO 
candidates sample, and in comparison with the results by \citet{Fla08}.

\begin{figure}
\centering
\includegraphics[width=9cm]{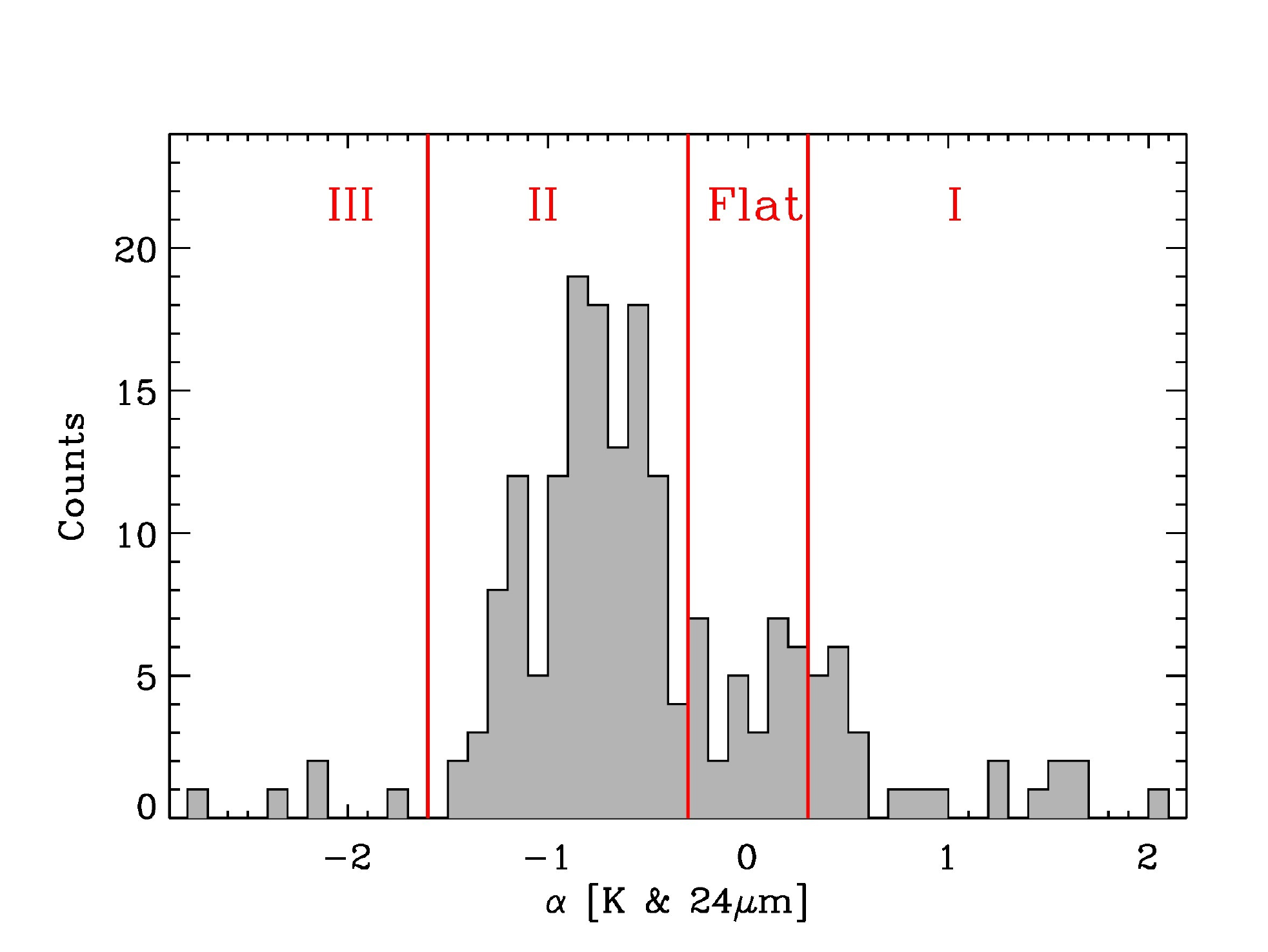}
\caption{$\alpha$-slope distribution of YSOc  in   L1630\,N. The vertical lines indicate the intervals 
defining the four Lada classes. The populations is largely dominated by Class~II objects.}
\label{fig_class}
\end{figure}

In order to investigate the disk properties of our YSO candidate sample, we adopted the Lada classification \citep{Lad84} 
based on the SED slope ($\alpha$) of the line joining the flux measurements at 2.2~$\mu$m (K-band)  and MIPS-24~$\mu$m. 
In particular, we used the Lada's class separation as extended by \citet{Gre94}, i.e.\ $ \alpha \ge 0.3$ for 
Class~I, $-0.3 \le \alpha < 0.3$ for flat-spectrum sources, $-1.6 \le \alpha < -0.3$ 
for Class II sources, and $\alpha < -1.6$ for Class III sources. We report in Table~\ref{tab_yso} the Lada class computed 
for each YSOc and give in Table~\ref{tab_class} the statistics for the entire YSOc sample in   L1630\,N. 
As shown in Figure~\ref{fig_class}, the dominant objects in   L1630\,N     are those of Class II (68\%), followed by 
flat-spectrum (16\%) and Class I (13\%) sources, with only a minority being Class III sources (3\%). 
The distribution of YSOs over class supports the young age estimated for this star-forming region. 
The ratio of the number of Class I and flat-spectrum sources to the number of Class II and Class III sources is 0.42, 
similar to the ratio measured in Serpens \citep{Har07} and other clouds of similar age surveyed by the {\it Spitzer} c2d 
survey \citep{Eva09}. 
The total observed fraction of objects with thick disks and/or envelope (Class I to II) for our sample is on the order of 97\%, 
while those with thin or no disk (Class III) represents only $\sim$3\% of the sample. The total disk fraction is 
considerably higher than the values derived in other regions of similar age \citep[e.g., in IC~348;][]{Lad06} and 
this would make L1630\,N a clear outlier with respect to 
the typical disk fraction vs. age trend \citep[e.g., see Fig.~1 and Fig.~4 by][respectively]{Hai01, Fed10}.
 
\begin{table}
\caption{Summary of Lada Classes in L1630\,N and estimated lifetime for each phase.}           
\label{tab_class}      
\centering                       
\begin{tabular}{cccc}      
\hline\hline               
Lada class &  n. of YSO candidates & lifetime  \\   
                    &                 &  (Myr)       \\   
\hline   
I         &    25  (13\%)	& 0.40 \\                     
Flat      &   30  (16\%) & 0.48   \\  
II        &   126   (68\%)  &  2$^\dag$  \\   
III       &   5   (3\%)    & --  \\   
\hline                                   
\end{tabular}
\\
$^\dag$ Assumed lifetime for the Class~II phase \citep{Eva09}.
\end{table}

Contamination from field stars, which could be as high as 25-30\%  (Sect.~\ref{vermin}) should not heavily 
affect the relative number of object with and without disk, because there is no reason to assume that this 
kind of contamination affects one Lada class more than the others. Background galaxies preferentially mimic 
the colors of YSOs with thick disks/envelope, but they may account for 2\% of our YSOc sample at most and 
can not justify the high number of Class I to II objects.

However, one may wonder whether our census might have missed a significant number of diskless YSOs.  
This would be possible because the c2d criteria select only IR excess objects.  
A direct way to investigate the number of diskless YSOs missed in our survey is to compare our results with 
those of deep X-ray observations, which are the most secure way to trace the population of class~III objects. 
By the merging of many {\em Chandra} pointed observations in a region centered between NGC\,2068 and HH24-26, 
\citet{principe14} obtained a very deep X-ray image with an equivalent exposure time of about 240$\times$10$^3$sec. 
In an area of about 0.12 square degrees they detected 52 X-ray sources, 32 of which can be identified with 
class~III or transition objects. Six of these objects have been recovered in the same area by our selection, 
but five were classified as Class~III and one as Class~II by our work. Considering that transition objects may mimic 
colors of Class~III objects, the conclusion is that our selection misses a factor of about 5 or 6 the actual 
number of diskless YSOs. This is in perfect agreement with the estimated number of missed diskless YSOs in 
the c2d surveys \citep[see Sect.~3.2 in][]{Eva09}. Hence, we are most likely missing 25-30 class~III 
YSOs (i.e.\ $\sim$20\%) in our sample (see Table~\ref{tab_class}), which is also consistent with the fraction 
of diskless YSOs missed by our criteria in the \citet{Fla08} and \citet{Fang09} samples 
(see Sect.~\ref{preioussurveys}). Certainly, in order to obtain a real full census of class~III objects in the area 
a much larger scale deep X-ray survey would be needed. But since such observations are currently not available
we use the above correction as a first order estimate, which leads to an expected fraction of class~III YSOs 
in L1630\,N predicted by our data of $\sim15$\%. This is still lower than expected on the basis of a 
cluster age of 1-2Myrs.

The surveys for PMS objects by \citet{Fla08} and \citet{Fang09} produced a similar result on the
high fraction of YSOs with disks and envelopes. Their surveys are rather complete in both space and flux and 
are based on different selection criteria, i.e., location on optical/near-IR color-magnitude diagram with 
respect to the  expected position of the main sequence at the cluster distance  and subsequent spectroscopic 
follow-up. \citet{Fla08} report a fraction of strong disk (Class I/II) of 66\% and a fraction of MIPS-weak disks 
of 16\%, in perfect agreement with our fraction of disk objects after applying the correction of missed 
class~III sources. They also find a fraction of 
IRAC-weak disks (Class~III) of 20$\pm$8\%, higher than our estimate but still lower than 
the fraction of Class~III YSOs found in regions of similar age. Similarly, \citet{Fang09} report a high 
fraction ($\sim$80\%) of disks in the region, although this value might also be biased by their selection 
which preferentially selects disk bearing young stars.

Alternatively, the substantial number of objects in younger SED classes is due to still ongoing 
star formation in L1630\,N, and correspondingly young age ($\le$\,1\,Myr) for the studied YSO 
samples. As we will see in Sect.~\ref{sec_sfr}, some studies \citep{Lad10,Hei10}  have 
revealed that, if most of the present-day mass measured for a given cloud lies below a certain gas surface density 
threshold, which was determined by \citet{Hei10} to $\sim129 \, \mathrm{M}_\odot\,\mathrm{pc}^2$ 
(corresponding to $\mathrm{A}_\mathrm{V}\approx8.6$~mag), 
a decrease in star formation 
could plausibly be caused by exhaustion of gas above such a threshold in surface density.
\citet{Spe11} demonstrated that this is the case for some clouds in the Lupus complex (Lup~V and VI), where 
only $\sim$1\% of the cloud mass lies above the threshold and, consistently, older SED classes (Class III) 
dominate the YSO population, while other Lupus clouds with 5 to 25\% of the cloud mass above the threshold 
are mainly populated by younger SED classes (Class I to II). The fraction of cloud mass above the threshold 
in L1630\,N is $\sim$35\% (Table~\ref{tab_SF}), i.e., even higher than in the most active star-forming region 
of Lupus (Lupus~III), and may explain its exceptionally high disk fraction.

Thus, although the actual value might be slightly lower, the result on the high disk fraction in L1630\,N seems 
to be real. The average disk fraction vs. age trend reported in the literature \citep[e.g.,][]{Fed10} suggests  
a median disk lifetime around 2-3~Myr, meaning that $\sim$50\% of the stars in a given population should have lost 
signatures of their disks after this time. However, several cases of clear outliers with respect to the average 
disk fraction vs.\ age trend have been reported in the literature. For example,  \citet{Alc08} found that only 
about 20\%-30\% of YSOs in Chamaeleon~II have lost their primordial disks in about 4 Myr (i.e., the average age 
for its members), and \citet{Sic06,Sic13} measured a disk fraction of $\gtrsim$50\% in the coeval cluster Trumpler~37.
Moreover,  studies in NGC~3603 \citep{Bec10} and the Magellanic Clouds \citep{Spe12,DeM11a,DeM11b} indicate that 
a considerable fraction of PMS stars still exhibit signatures of accretion from a circumstellar disk at ages 
as old as 10~Myr. It is still under debate whether these differences from region to region are due to residual 
incompleteness effects of different surveys, limitations of the adopted selection methods, etc., or to the 
specific properties of the given star-forming environment (such as metallicity, presence of strong UV radiation 
fields, multiplicity, crowding, etc.), which may strongly affect disk evolution \citep[e.,g.,][]{Hol00,Lin07,Dul05,Joh09,Dae13}.

\citet{Eva09} derived the half-life for each of the Lada classes from the combined analysis of the Spitzer c2d data set. 
According to this study, the half-life for Class II sources is $\sim$2~Myr. If star formation has been continuous over 
a period longer than the age of Class II sources, the lifetime for each phase can be estimated by taking the ratio 
of number counts in each class with respect to Class II counts and multiplying by the lifetime for Class~II. 
According to the statistics of Lada classes in  L1630\,N, we estimate a lifetime of 0.4 and 0.48~Myr for the 
Class~I and Flat-spectrum phase, respectively (Table~\ref{tab_class}). These values agree with the lifetimes 
derived by \citet{Eva09} by averaging all c2d clouds.

\addtolength{\tabcolsep}{0.5pt}     
\begin{table*}
\caption{Results of the clustering analysis and properties of star formation in sub-structures in Lynds 1630\,N}           
\label{tab_clustering}      
\centering                       
\begin{tabular}{l|l|cccc|lllccc}      
\hline\hline               
Region      &  \#YSOs     & \multicolumn{4}{c|}{\# IR Class}                     & Mass        &  $\langle A_V \rangle$ &  SFE    & Area        & Mass/Area  ($\Sigma_{gas}$) & SFR/Area  ($\Sigma_{SFR}$) 	  \\   
            &             &       I   &   Flat &        II &         III &  (M$_\odot$) &          (mag.)       &  \%     & (pc$^{2}$) & (M$_\odot$ pc$^{-2}$)       & (M$_\odot$ Myr$^{-1}$ pc$^{-2}$) \\   
\hline   
Extended    &    15       &     4     &   3    &     6     &      2      &     1840	&	   0.3  	 &  0.4    &     ...          &  ...  &  ...   \\   
Lynds 1630\,N  &    171      &     21    &   27   &     120   &      3      &     2050	&	   4.8  	 &  4.0    &     20.2         &  102  &  2.0   \\   
...NGC 2071 &    52       &     5     &   12   &     34    &      1      &     400	&	   8.8  	 &  3.8    &     2.17         &  180  &  6.0   \\   
...NGC 2068 &    45       &     4     &   5    &     36    &      0      &     243	&	   5.5  	 &  5.3    &     2.12         &  115  &  5.4   \\   
...HH24-26  &    14       &     6     &   2    &     5     &      1      &     124	&	   7.7  	 &  3.3    &     0.77         &  161  &  4.6   \\   

\hline                                   
\end{tabular}
\end{table*}

\section{Spatial Distribution and Clustering \label{sec_spa_dist}}

The spatial distribution of the different classes of objects in L1630\,N is shown in Figure~\ref{spa_distr}  over-plotted on the VISTA 
extinction map, derived as explained in Sect.~\ref{sec_ext}. The Class I and Flat sources  coincide or are located close to the sites of 
highest extinction, as observed in all young clusters still associated with the residual parental clouds. 

NGC~2068 and  NGC~2071 are the most prominent star-forming clusters in the L1630\,N  molecular cloud. 
The VISTA extinction map shows, in a consistent way, two extinction peaks corresponding to the approximate center of NGC~2068 
(R.A.$\approx$86.65~deg, Dec.$\approx$0.1~deg) and  NGC~2071 (R.A.$\approx$86.75~deg, Dec.$\approx$0.35~deg). 
It is also evident that, beside these two peaks, there is an additional extinction peak centered at R.A.$\approx$86.55~deg and  
Dec.$\approx$-0.15~deg. This third peak corresponds to the HH~24-26 group of Herbig-Haro objects, and is very close to V1647~Ori 
\citep[see Fig.~7 by][]{Gib08}, a low-luminosity protostar, perhaps in a transition phase from Class I to Class II which underwent 
a strong outburst in 2004 \citep{Bri04}. We also observe a very clear concentration of Class I and Flat sources around this peak, 
confirming that the region is one of the several small centers of star formation in L1630. Another argument for still 
actively ongoing star formation, is a notable concentration of presumably very young protostars in the region \citep[see][]{stutz13}.

\begin{figure}
\centering
\includegraphics[width=9.5cm]{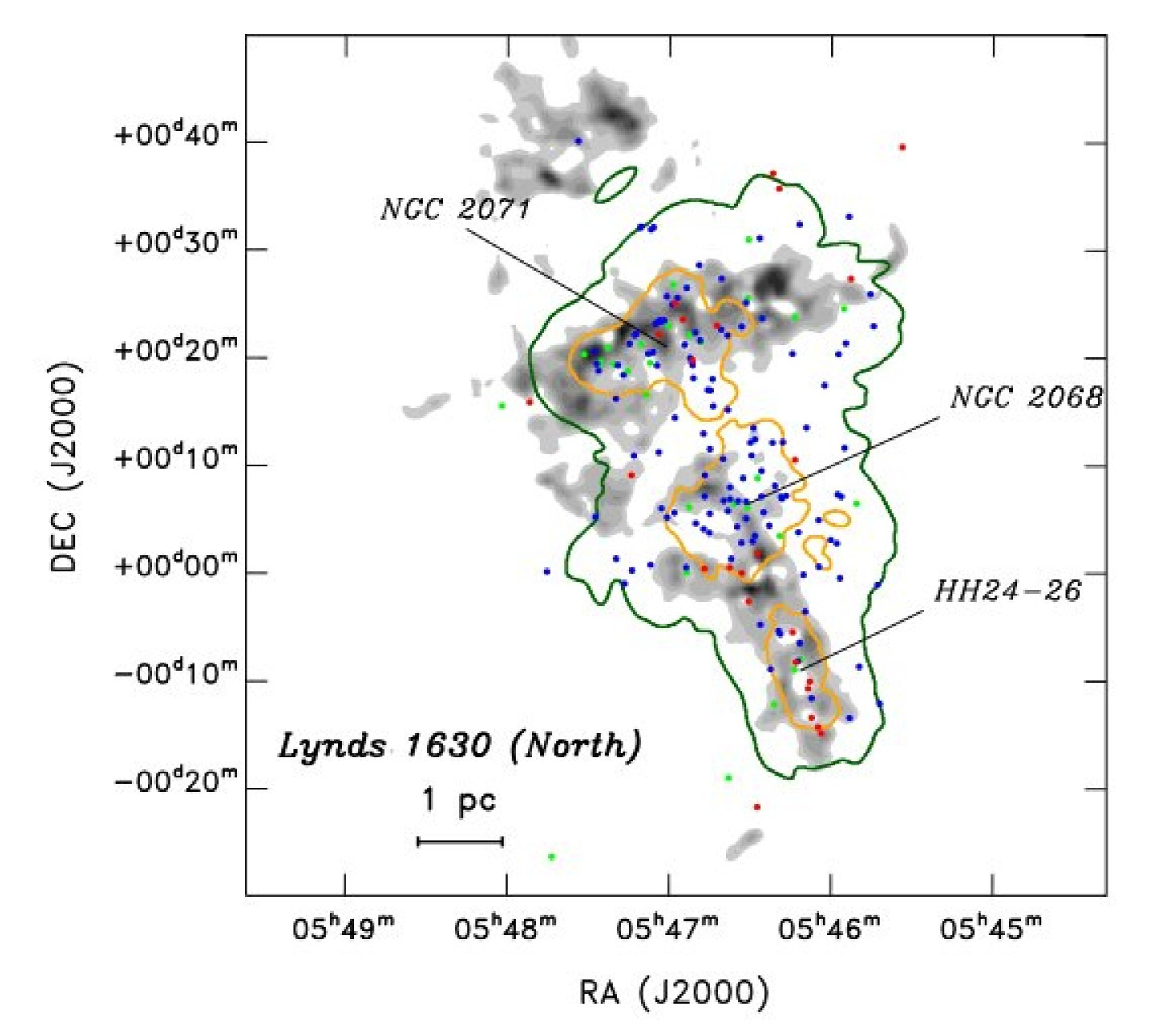}
\caption{Distribution of young stellar objects plotted over the extinction map (in gray) of the region. 
        The green and orange contours correspond to volume densities of 1~M$_\odot$ pc$^{-3}$
	and 10~M$_\odot$ pc$^{-3}$, respectively. Class~I sources are shown in red, Flat spectrum 
	sources in green and Class~II sources in blue. }
\label{fig_clustering}
\end{figure}

\citet{Lad03} suggested that a cluster should be a group of some 35 members with a total mass density larger 
than $1 \, \mathrm{M}_\odot\,\mathrm{pc}^{-3}$.
To compare with this criterion and in order to assess the sub-clustering structures in L~1630, it is important to examine the distribution 
of YSOc in a quantitative and uniform way. 
We calculated the volume density of YSOs, based on their Lada classes and position, using a nearest 
neighbor algorithm similar to the one applied by \citet[][]{Gut05} and implemented by \citet{Jor08} for the c2d clouds. The calculations assume 
that the distribution of sources is locally spherical. We applied the algorithm to the whole sample of 186 YSOs in Table~\ref{tab_yso}. 
The overall results, reported in Table~\ref{tab_clustering}, are shown in Figure~\ref{fig_clustering}.

To define a cluster the c2d surveys adopted the tighter level of 25 times the \citet{Lad03} criterion (i.e., a mass density of 
$25 \, \mathrm{M}_\odot\,\mathrm{pc}^{-3}$),
which normally provides the already established cluster and group boundaries. Within the c2d surveys, ''clusters'' are regions with more than 
35 YSOs within a given volume density contour and ''groups'' are regions with less. The lowest number of YSOs that is considered to 
constitute a separate entity is 5. ''Clusters'' and ''groups'' can be either ''tight'' or ''loose'' depending on whether their volume 
densities, $\rho$, are higher than 25 or 1~M$_\odot$ pc$^{-3}$ (corresponding to 50 and 2 YSOs pc$^{-3}$), respectively, assuming an 
average YSO mass of 0.5M$_\odot$. 
As noted in the c2d papers, these criteria are useful as a way of making direct comparison between regions within clouds and across different 
clouds, but being empirical they should not be used as evidence for discussions on whether the star formation process is hierarchical or not.

In L1630\,N we identify structures with two levels of volume densities, structures with a density larger than 1~M$_\odot$ pc$^{-3}$ 
\citep[similar to the criterion for a cluster by][]{Lad03} or structures with a density as high as 10~M$_\odot$ pc$^{-3}$. 
Note, however, the latter is lower than what was applied 
for "tight" associations in the c2d surveys. Thus, according to the c2d criteria, the loose clusters identified in this region are
L1630\,N as a whole with a density of 1~M$_\odot$ pc$^{-3}$ and NGC\,2071 and NGC\,2068 with a density of 10~M$_\odot$ pc$^{-3}$.
The HH24-26 entity can be defined as a loose group with a density of 10~M$_\odot$ pc$^{-3}$.

\section{Overall results on star formation in L1630\,N }

In this section we analyze and discuss the cloud properties of   L1630\,N and the global properties of star 
formation in this region on the basis of the VISTA observations  and our YSOc sample. Our results are summarized 
in Table~\ref{tab_SF}.

\subsection{Extinction map, cloud mass and surface density  \label{sec_ext}} 

An extinction map, with a resolution of 30$\arcsec$, was constructed for the entire area of Orion observed in the 
VISTA SV mini-survey (Lombardi et al., {\it in prep.}).  The technique used is  optimized to produce highly accurate 
extinction maps from multi-band near-IR photometric data as outlined in \citet[][NICER]{Lom01} and \citet[][NICEST]{Lom09}.
The method is the natural generalization of the near-infrared color excess (NICE) method of \citet{Lad94} and produces 
significantly less noisy and, hence, more accurate extinction maps taking advantage of all bands available. Applications 
of this technique to 2MASS data has shown an improvement with respect to the standard NICE algorithm 
of a factor 2 on the noise variance \citep{Lom01}. 
We further compared our extinction map values with the spectroscopic values inferred by \citet{Fla08} for a sample 
of 67 previously known PMS stars in the cloud and found a good agreement (within 2-3~visual mag), 
but also an apparent shift of A$_V = 3.4$mag (i.e. the extinction map gave a consistently higher A$_V$ than what 
was determined from spectroscopy). This has been observed for other regions as well and can be explained 
by the fact that the extinction map measures the extinction through the {\it whole} cloud, while the PMS stars 
are located {\it in} the cloud. 

According to this extinction map, the extinction in the   L1630\,N  cloud (tile~no.12) is typically low ($A_{V}\approx$1~mag), 
with peaks up to A$_V \approx$20~mag occurring close to the following locations:  R.A.=86.75~deg \& Dec.=0.35~deg 
(i.e., the center of NGC~2071), R.A.=86.65~deg \& Dec.=0.1~deg (i.e., the center of NGC~2068) and R.A.=86.55~deg \& Dec.=-0.15, 
where a group of YSOc has been identified by us (Sect.~\ref{sec_spa_dist}).

Using the extinction map and assuming a distance of 400~pc, we also estimated the cloud 
mass for the  L1630\,N  complex. To this aim we used the relationship between gas 
surface density, $\Sigma_{gas}$,  and extinction by \citet[][]{Hei10}, i.e.
$\Sigma_{gas}~=~15~\times~{\rm A_V}$\,M$_\odot$/pc$^2$. We estimated that the total cloud 
mass for $A_V\ge2$ mag is about 3865\,M$_\odot$\footnote{Note 
that the total mass of gas in dense cores in L1630\,N has been estimated to be $\sim$2000\,M$_\odot$
\citep[see Sect.~3.1 in][]{Gib08}.}; considering that the area where A$_{\rm V} \ge$2~mag extends over 
$\sim$39~pc$^2$, the cloud column density is about $\sim$100~M$_\odot$ pc$^{-2}$. The NICE and NICER methods 
provide an intrinsic error of about 0.5\,mag, on the average. Assuming this value as the 
intrinsic error of our VISTA extinction map, and an error on the distance to the cloud of about
10\% \citep[see Sect.~2 in][]{Gib08}, we estimated that the cloud mass and column density of L1630\,N  
can be safely placed in the ranges 3200$-$4600~M$_\odot$ and 81$-$116~M$_\odot$ pc$^{-2}$, respectively. 

We stress that in $\sim$35\% of the cloud area where $A_{\rm V} \ge 2$\,mag the extinction is above 8.6\,mag. 
As concluded in \citet[][]{Hei10} the latter value for the extinction sets an 
important threshold at which the gas surface density is linearly proportional to the surface density 
of the star formation rate.
 
\subsection{Star formation efficiency \label{sfe}}

We derive the global star formation efficiency (SFE) in L1630\,N as:
\begin{equation}
SFE=\frac{M_{stars}}{M_{cloud}+M_{stars}} 
\end{equation}
where M$_{cloud}$ is the cloud mass derived in Sect.~\ref{sec_ext}, and M$_{star}$ is the total mass converted into stars. 
M$_{star}$ is derived from the number of YSOc identified in our study, i.e.\ without applying any corrections
for missed diskless star, following the procedure applied by the {\it Spitzer}-c2d/GB surveys.  As already pointed 
out by \citet[][]{Eva09}, this implicates that the star formation efficiencies and rates over the whole 
star forming cloud's lifetime could be higher. 
For L1630\,N we estimated M$_{star}$ to be $\sim$93~M$_\odot$, assuming an average YSO mass of 0.5~M$_\odot$ 
consistent with the peak observed in the KLF (Fig.~\ref{KLF}), and with the assumption made in all clouds observed 
by the {\it Spitzer}-c2d/GB surveys, which we use for comparison. 

We find that the  overall SFE in L1630\,N ranges between 2\% and 2.8\% and is $\sim 4.0$\% on average for the
L1630\,N clusters (Tab.~\ref{tab_clustering}), which is overall consistent
with the typical values measured for Orion~A and B 
\citep[][and references therein]{Fed13}, for all c2d clouds \citep[see Table~4 by][]{Eva09} and, more in general, for the 
majority of star-forming regions in the Galaxy \citep[e.g.,][]{Fed13}. However, we notice that the SFE in L1630\,N is lower 
than measured in the Orion bright-rimmed clouds \citep[e.g., 5 to 10\%;][]{Lee05} and, in particular, lower than measured 
in sub-clusters in the southern region L1630\,S.
For the populous cluster NGC~2024, \citet{Lad97} determined a high SFE of $\sim$30\%, although
based on CS observations that trace only the highest density gas, and hence providing a lower cloud mass than
the mass obtained by us via the extinction map method. It appears that the CS measurements underestimate 
the total cloud mass by a factor of 3-4 as compared to the cloud mass derived from the extinction map,
such that the true SFE for NGC~2024 is likely more like $\sim10$\%. 
In comparison, the SFE for the sub-clusters identified in L1630\,N is very low (see Table~\ref{tab_clustering}).
We recall that it is expected that star formation may be more efficient in localized, compressed regions, where triggered star 
formation might play a role, but perhaps not so in the entire cloud \citep{Lee05}. 
The different SFE measured between the northern and the southern regions of L1630 might indicate that two different 
star-formation mechanisms currently dominate in the two regions of this cloud.

\addtolength{\tabcolsep}{-4pt}     
\begin{table}
\small
\caption{Overall properties of the star formation in L1630\,N    .}           
\label{tab_SF}      
\centering                       
\begin{tabular}{lccc}      
\hline\hline               
Property &  Value & Uncertainty Range  & Unit \\    
\hline   
peak of the KLM  &  0.5 & 0.3-0.7 & M$_\odot$ \\   
cloud area (A$_V\ge$2~mag) & 39 & -- &   pc$^2$   \\
cloud mass  (A$_V\ge$2~mag) &  3865  &  3200-4600 & M$_\odot$  \\ 
cloud density ($\Sigma_{gas}$) & 98 & 81-116 &   $\mathrm{M}_\odot\,\mathrm{pc}^{-2}$  \\                 
Fraction of cloud above $\Sigma_{th}$ & 0.35 & -- &  \\
N. YSOc/Area & 5 & -- & pc$^{-2}$ \\
SFE    &   2.35 & 2-2.8 & percent \\  
SFR      & 75  & 47-103 &   M$_\odot$/Myr \\   
SFR/Area  ($\Sigma_{SFR}$)  &  1.9 & 1.2-2.6 &  M$_\odot$ Myr$^{-1}$  pc$^{-2}$  \\   
\hline                                   
\end{tabular}
\end{table}


\subsection {Star formation rate and star formation density \label{sec_sfr}} 

We derive the star formation rate (SFR) in L1630\,N as:
\begin{equation}
SFR=\frac{M_{stars}}{Age} 
\end{equation}
where M$_{star}$ is the total mass converted into stars (equal to $\sim$93~M$_\odot$, Sect.~\ref{sfe}) and $Age$ is the average 
age of the YSO population. The latter has some uncertainty and therefore determines the possible
range for the resulting SFR.  A median age of 2\,Myr was determined by \citet{Fla08} for their optical spectroscopy 
sample in NGC~2068/71. However, this could be an overestimate for the L1630\,N YSO sample in 
our work, which does include the large majority of the FM08 objects but also 
include highly embedded, i.e.\ potentially much younger, sources. Note that \citet{Fang09} estimate a median
age of 0.9\,Myr for their YSO sample of L1630\,N, of which 60\% are included in our survey.


Taking the age uncertainty into account, we find that the  L1630\,N  cloud is turning some 
$75\pm28$~M$_\odot$ into YSOs every Myr. This SFR is in agreement with the SFR 
measured for the c2d/GB clouds \citep[see Table~3 by][]{Eva09}, with the exception of Cha~II, where the  SFR seems to be very 
low \citep{Alc08}. However, the SFR in L1630\,N appears lower than the average SFR measured for the overall Orion A and B molecular 
clouds \citep[150-700~M$_\odot$/Myr; see Table 2 by][]{Lad10} and the local SFR measured for the ONC \citep{Lad96} and Trapezium 
\citep{Pal99,Lad03}. We note, however, that large SFR variations are observed among the Orion sub-regions; for example, the SFR 
in L1630 (to which L1630\,N belongs) is known to be a factor of 2 to 7 lower than observed in the nearby L1641 \citep{Mey08}. 

These variations can ben be reconciled if instead we consider the SFR per unit area ($\Sigma_{SFR}$), i.e., the density of star 
formation. It has been confirmed that $\Sigma_{SFR}$ is linearly proportional to the cloud gas surface density 
($\Sigma_{gas}$), above an extinction threshold of A$_\mathrm{V} \approx$8.6~mag  \citep{Hei10}, corresponding to a gas 
density threshold ($\Sigma_{th}$) of $\sim$129~M$_\odot$\,pc$^{-2}$. We measure for L1630\,N a $\Sigma_{SFR}$ of 
1.9~M$_\odot$ Myr$^{-1}$  pc$^{-2}$ and $\Sigma_{gas}$ of $\sim$98~ M$_\odot$\,pc$^{-2}$, and these values are in excellent 
agreement with previous observations of galactic star-forming activity \citep{Hei10}. Note also that, as mentioned in 
the previous section, about 35\% of the cloud has A$_\mathrm{V} >$8.6~mag. Thus, more than one-third of the cloud has 
a $\Sigma_{gas}$  above $\Sigma_{th}$.
At the level of the sub-structures identified in the clustering analysis, the values of $\Sigma_{SFR}$  and  $\Sigma_{gas}$
are slightly higher (see Table~\ref{tab_clustering}) but still well within the range for galactic star-forming 
regions \citep{Hei10}.

The linear correlation between the rate of star formation and the amount of dense gas in molecular clouds, confirmed by 
all c2d/GB clouds \citep{Hei10},  all nearby molecular clouds \citep{Lad10}, galactic massive dense clumps \citep{Wu10}, 
the youngest and still embedded Class I and Flat-spectrum YSOs in the Galaxy \citep{Hei10},  and also consistent 
with the results for several nearby molecular clouds \citep{Gut11},
lies above the extragalactic SFR-gas relations \citep[e.g., Kennicutt-Schmidt law;][]{Ken98} up to a factor of 17 to 54 \citep{Hei10}.
Moreover, the extragalactic SFR-gas relation is not linear, because $\Sigma_{SFR}$ scales as $\Sigma^{1.4} _{gas}$.
Several contributing factors to this difference have been identified so far \citep{Hei10}: 
i) much of $\Sigma_{gas}$ is below $\Sigma_{th}$ in extragalactic studies, which average over large scales and include both 
star-forming gas and gas that is not dense enough to form stars, ii) using ${}^{12}\mathrm{CO}$ or ${}^{13}\mathrm{CO}$  to measure the H$_2$ in 
galaxies gives systematically lower $\Sigma_{gas}$ than Galactic A$_\mathrm{V}$ measurements, as the one we used, by a factor up to 30\%.
Indeed,  power-law indices between 0.8 and 1.6 have been found for the extragalactic SFR-gas relation \citep[e.g.,][]{Ken07,Big08,Kru09}, 
depending on the survey spatial resolution and the adopted tracer. These overall results suggest that the key to obtaining a 
predictive understanding of the star formation rates in molecular clouds and galaxies is to understand those physical factors 
which give rise to the dense components of these clouds \citep{Lad10}.

\section{Summary and conclusions}

Based on the VISTA Orion mini-survey, complemented  with Spitzer observations, we have performed a 
study of the YSO population and star formation in the L1630\,N cloud. The c2d multi-color criteria 
selected 186 YSOs in the area of about 1.5~square degree in L1630\,N. The census is $\sim$95\% complete 
down to M$_{\star}\sim$0.02 M$_{\odot}$. We have investigated both the YSOs with infrared excess selected 
according to the c2d criteria, as well as the other YSOs cloud members and candidates from the previous 
surveys. Spectroscopic follow-ups, published in the literature, confirm the YSO nature of most of the 
selected candidates, supporting the reliability of the selection criteria. 

The K-band luminosity function of L1630\,N shows a broad peak between 10.5 and 12\,mag., i.e. 0.3-0.7~M$_{\odot}$, 
but steadily declines to the hydrogen-burning limit at K$_S \approx$13\,mag. We predict a fraction of 28\% young 
substellar objects, but we note that the expected contamination from field stars mimicking the colors of BDs is 
on the order of 50\%, while in the stellar regime it is about 20\%. Thus, the actual fraction of substellar objects
in L1630\,N may be $\sim$20\%. The K-band luminosity function of L1630\,N is remarkably similar to that of the
Trapezium cluster.

The analysis of the SEDs shape shows that the L1630\,N population is dominated by objects with active accretion, 
with only a minority being systems with passive disks. The disk/envelope fraction in the region of $\sim$85\% is 
high in comparison with other star formation regions of similar age. 
The fraction of the Class~I and Flat-spectrum sources 
(13\% and 16\%, respectively) in L1630\,N and their respective phase lifetime (0.4 and 0.48 Myr, respectively) 
are consistent with the results for the c2d clouds \citep[][]{Eva09}. 
 
We studied the spatial distribution and volume density of the 186 YSOs with the following results: 
we identify structures with volume densities higher than 1~M$_\odot$ pc$^{-3}$ or 10~M$_\odot$ pc$^{-3}$. 
The loose clusters identified are L1630\,N as a whole with a density of 1~M$_\odot$ pc$^{-3}$ and NGC\,2071 
and NGC\,2068 with a density of 10~M$_\odot$ pc$^{-3}$. The HH24-26 entity can be defined as a loose group 
with a density of 10~M$_\odot$ pc$^{-3}$.

The cloud mass determined from the VISTA extinction map is on the order of 3865 $M_{\odot}$ and the SFE 
of 2-2.8\% is similar to previous estimates for the Orion A and B clouds and for the c2d clouds,  but is much 
lower than the SFE measured in sub-clusters in the southern region L1630\,S. The SFE of the sub-clusters in  
L1630\,N is also comparably low. The different SFE in the northern and southern regions of L1630 might 
suggest different star formation mechanisms. The SFR is similar to that of the c2d clouds \citep[][]{Eva09}; 
we find that L1630\,N is turning some 75~$M_{\odot}$ into YSOs every Myr. This is, however, lower than the 
average value measured for the Orion A and B clouds and the local SFR for the ONC and the Trapezium, but
large variations of the SFR among the Orion sub-groupings are observed. Such variations disappear
when considering the density of star formation $\Sigma_{SFR}$. The density of star formation 
$\sim$ 2\,M$_\odot$\,Myr$^{-1}$\,pc$^{-2}$ and the gas surface density $\sim$98\,M$_\odot$\,pc$^{-2}$ in
L1630\,N are in excellent agreement with previous determinations of galactic star forming activity.
At the level of the sub-clusters in L1630\,N  these quantities are also similar to those in the 
sub-clusters in other galactic star forming regions.
More than one-third of the cloud in L1630\,N has a gas surface density above $\Sigma_{th}\sim$129~M$_\odot$/pc$^2$. 
This may indicate that star formation in L1630\,N is still on-going, which may explain the 
exceptionally high disk/envelope fraction in the region. The latter, however, needs to be confirmed in the future
with deep observations tracing the complete population of young diskless sources.
 
\begin{acknowledgements}

We thank the anonymous referee for valuable comments which further improved the 
clarity of the paper. 
JMA acknowledges financial support from INAF under the program PRIN2013
"Disk jets and the dawn of planets". This research has made use of the SIMBAD 
database operated at CDS, Strasbourg, France. It also makes use of data products 
from the Two Micron All Sky Survey, which is a joint project of the University 
of Massachusetts and the Infrared Processing and Analysis Center/California 
Institute of Technology, funded by the National Aeronautics and Space 
Administration and the National Science Foundation. We greatly appreciate the 
work done by the UK-based VISTA consortium who built and commissioned the 
VISTA telescope and camera.
    
\end{acknowledgements}

\begin{appendix} 
\section{Theoretical isochrones in the VISTA photometric system \label{isocr_vista}}

Theoretical isochrones for low-mass stars and BDs down to 0.001~$M_{\odot}$ are provided by \citet{Bar98} and \citet{Cha00} in the Cousins photometric system \citep{Bes90}, 
and are the most commonly used for very low-mass stellar population studies. In particular, they are extensively used to select PMS star and young BD candidates on the basis of color-magnitude diagrams (CMDs).
Since the transmission curves of the VISTA filters are very different from the Cousins ones, we transformed these isochrones into the specific VISTA photometric system. 
In this way, we make available to the community a valuable tool to be used in extensive VISTA-based searches for very low-mass stars and BDs in other stars forming regions.
 
\begin{figure}
\centering
\includegraphics[width=7cm]{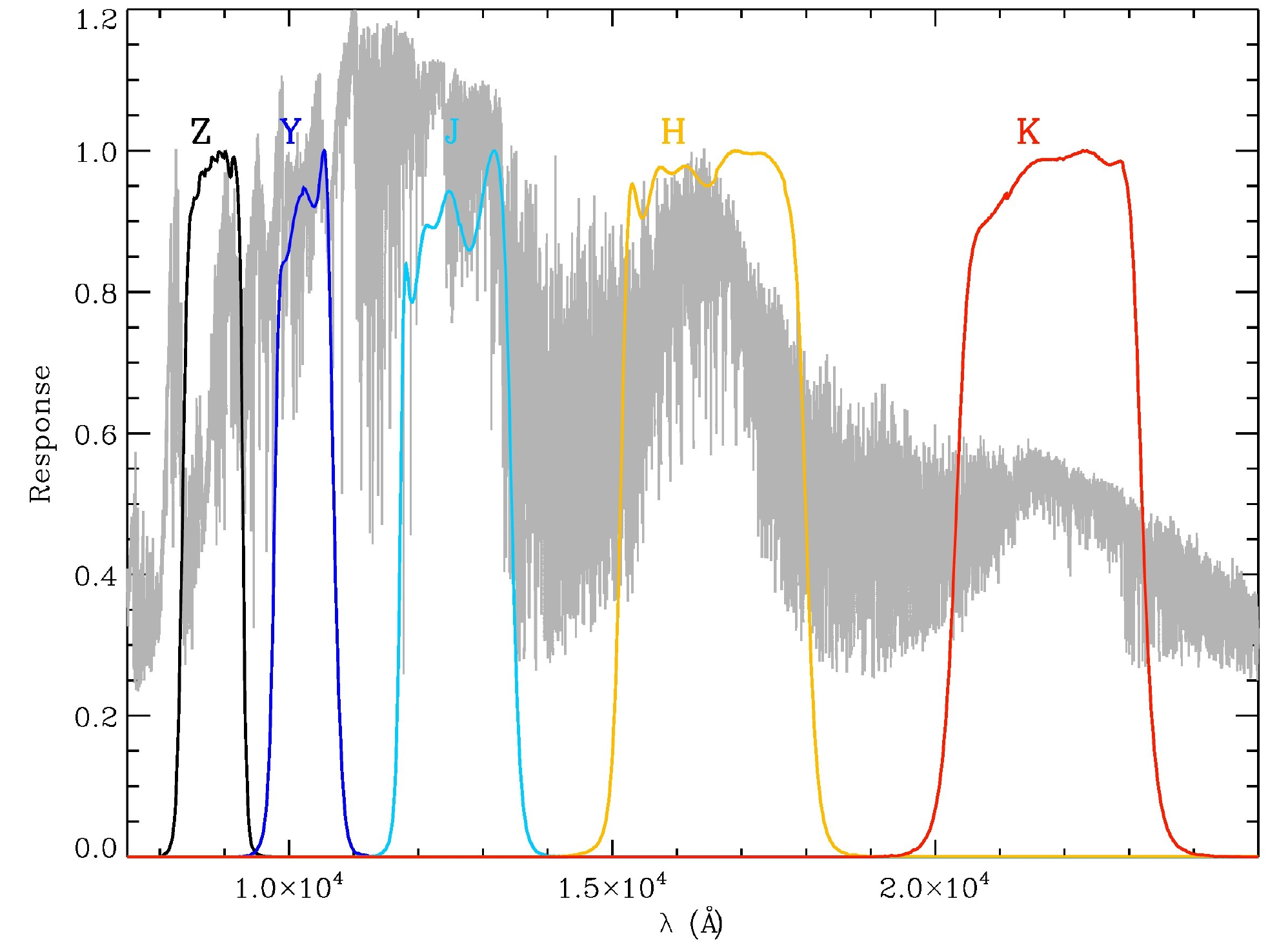}
\includegraphics[width=7cm]{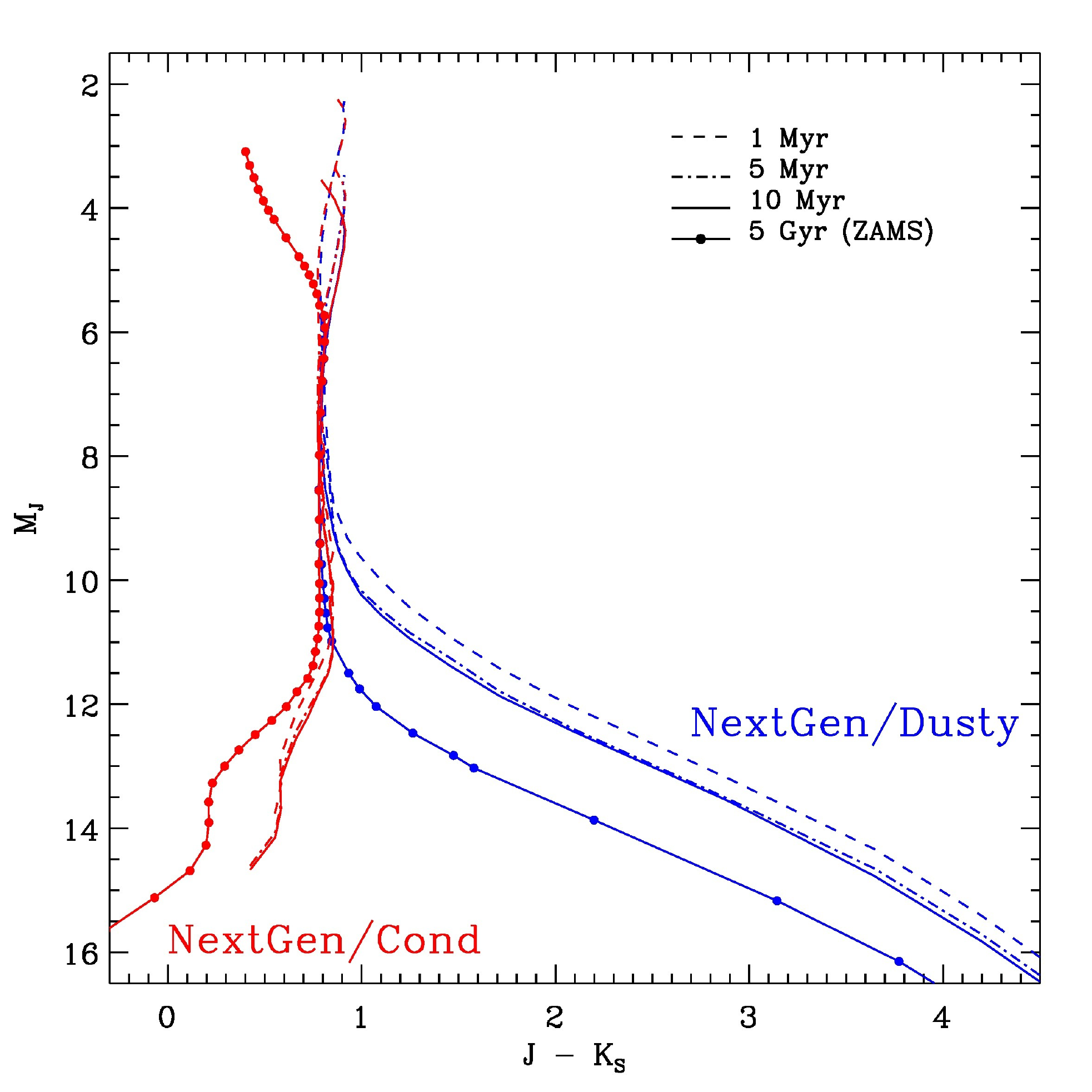}
\caption{
{\bf Upper panel:}  The $ZYJHK_S$ VISTA transmission bands. An example of a normalized NextGen spectrum \citep{Hau99} with $T_{\rm eff}=2500$ K is over-plotted. 
{\bf Lower panel:} Theoretical  $M_J$ vs. $J-K_S$ diagram. The isochrones, shifted to the distance of 10~pc (i.e., absolute magnitudes), are in the VISTA photometric system. 
Different curve-styles correspond to different ages, as indicated in the legend. We plot in blue the isochrones computed using the NextGen/AMES-Dusty atmospheric models, 
and in red those computed using the NextGen/AMES-Cond models. 
}
\label{fig_iso_vista}
\end{figure}

The procedure adopted to perform the conversion  of the evolutionary models  from one system to another has been already described in detail in  \citet{Spe07} (appendix~B). 
The expected flux ($f_{\Delta \lambda}$) at the stellar surface in the VISTA pass-bands were determined by integrating the synthetic low-resolution spectra for low-mass stars by \citet{Hau99}, 
calculated with their NextGen model-atmosphere code, under the  filter transmission curves\footnote{Available at http://apm49.ast.cam.ac.uk/surveys\-projects/vista/technical/filter\-set}
(see Fig.\ A.1, upper panel). 
For simulating very cool objects (i.e. $T_{\rm eff}<3000$ K) we used the AMES-Dusty and AMES-Cond atmosphere models by \citet{All01}, which take into account the formation of condensed 
species significantly modifying the atmospheric structure \footnote{While in the AMES-Dusty models the condensed species are included both in the equation of state and in the opacity, 
taking into account dust scattering and absorption, in the AMES-Cond models the opacity of these condensates is ignored, in order to mimic a rapid gravitational settling of all grains below the photosphere}. The expected flux
$f_{\Delta \lambda}$ was then converted to absolute magnitudes using the following equation:
 
 \begin{equation}
m_{\Delta \lambda} = - 2.5 \cdot \log_{10} \left(f_{\Delta \lambda} \cdot \frac{R_\star}{d}\right) + C_{\Delta \lambda}
\end{equation}
 
 where  d=10~pc, $R_\star$ is the  stellar radius expected for PMS objects and computed from the theoretical PMS evolutionary 
 tracks by \citet{Bar98} for low-mass stars and those by \citet{Cha00} for sub-stellar objects (i.e. $M\la 0.1~ M_{\odot}$), 
 and  $C_{\Delta \lambda}$ is the absolute calibration constant of the VISTA photometric system, 
 tied to the Earth flux of an A0-type star with magnitude V=0 \footnote{http://apm49.ast.cam.ac.uk/surveys\-projects/vista/technical/photometric\-properties}.
In Figure~\ref{fig_iso_vista} (lower panel) we show, as an example, the theoretical 1, 5 and 10~Myr isochrones and the ZAMS (5~Gyrs) on the $M_J$ vs. $J-K_S$ CMD and make them publicly available  in Table~\ref{tab_iso_vista}, where we also give isochrones for 50~Myr and 100Myr.

\tiny
\begin{longtable}{cc|ccccc|ccccc}      
\caption{ \label{tab_iso_vista} Theoretical isochrones from  \citet{Bar98} and \citet{Cha00} converted into the VISTA photometric system. }      \\                 
\hline\hline     
    Mass                     & T$_{eff}$  &  $Z$ & $Y$ & $J$ & $H$ & $K_S$ & $Z$ & $Y$ & $J$ & $H$ & $K_S$ \\   
  (M$_\odot$)          &      (K)         &         &          &        &           &              &         &          &        &         &                 \\   
\hline           
\endfirsthead
\hline\hline 
    Mass                     & T$_{eff}$  &  $Z$ & $Y$ & $J$ & $H$ & $K_S$ & $Z$ & $Y$ & $J$ & $H$ & $K_S$ \\   
  (M$_\odot$)       &      (K)         &         &          &        &           &              &         &          &        &         &                 \\   
\hline            
\endhead
\hline
\multicolumn{12}{l}{{Continued on Next Page\ldots}}
\endfoot
\hline \hline
\endlastfoot    
\multicolumn{12}{c}{NextGen/AMES-Dusty  \hspace{3.5cm} NextGen/AMES-Cond} \\ 
\hline
\multicolumn{12}{c}{1~Myrs} \\ 
\hline						  
   0.002    &      1300  &    20.899   &  18.949   &  16.523   &  13.923   &  11.824   &    15.192   &  13.621   &  12.496   &  12.326    & 11.875	   \\ 
   0.002    &      1400  &    19.306   &  17.576   &  15.410   &  13.066   &  11.211   &    14.251   &  13.354   &  12.133   &  11.904    & 11.470	   \\ 
   0.003    &      1500  &    17.958   &  16.376   &  14.336   &  12.237   &  10.691   &    13.900   &  13.042   &  11.777   &  11.516    & 11.055	   \\ 
   0.003    &      1600  &    16.643   &  15.123   &  13.181   &  11.525   &  10.300   &    13.945   &  12.610   &  11.426   &  11.147    & 10.646	   \\ 
   0.004    &      1700  &    15.110   &  13.702   &  12.031   &  10.881   &   9.949   &    13.754   &  12.172   &  11.105   &  10.812    & 10.278	   \\ 
   0.004    &      1800  &    14.359   &  12.961   &  11.422   &  10.511   &   9.710   &    13.320   &  11.773   &  10.788   &  10.490    &  9.941	   \\
\hline      			     	    	     	     	 	     		    
\end{longtable}

\tiny
\setlength\LTleft{-7cm}          
\setlength\LTright{10pt}    
\addtolength{\tabcolsep}{0pt}     

\begin{landscape}
\begin{longtable}{cccccccccccccc}
 \caption{ \label{tab_yso}  VISTA and {\it Spitzer} observed magnitudes for the YSO candidates in L1630\,N.  When the photometric errors are not reported, 
 the corresponding magnitude is an upper limit to the actual brightness of the object. The ``--'' symbol indicates that the object has not been detected in 
 the given VISTA pass-band. The Lada classes marked with ``?'' are uncertain because the corresponding K$_S$ magnitude is uncertain (i.e.\ photometric contamination, 
 large photmetric errors, object truncated or partially saturated, etc.). Column 3 specifies whether the object has been previously identified by \citet{Fla08} (F08) 
 and/or \citet{Fang09} (FA09) .}  \\         
\hline\hline               
ID & Name/Position & Previous       & Lada   &  $Z$ & $Y$ & $J$ & $H$ & $K_S$ & IRAC                & IRAC            & IRAC              & IRAC        & MIPS  \\   
    & (VISTAomsJ)    & Identification & Class &          &          &        &         &              &   3.6$\mu$m     & 4.5$\mu$m & 5.8$\mu$m    & 8$\mu$m &  24$\mu$m \\ 
\hline          
\endfirsthead
\hline\hline  
ID & Name/Position & Previous       & Lada   &  $Z$ & $Y$ & $J$ & $H$ & $K_S$ & IRAC                & IRAC            & IRAC              & IRAC        & MIPS  \\   
    & (VISTAomsJ)    & Identification & Class &         &          &        &         &              &   3.6$\mu$m     & 4.5$\mu$m & 5.8$\mu$m    & 8$\mu$m &  24$\mu$m \\ 
\hline            
\endhead

\hline
\multicolumn{14}{l}{{Continued on Next Page\ldots}}
\endfoot

\hline \hline
\endlastfoot    
              1  &      054803.999+000341.11        &            &    III?   & 10.635    & 10.153    &  8.723    &  8.467    &  7.167    &  6.526 $\pm$  0.002 &  6.559 $\pm$  0.001 &  6.353 $\pm$  0.001 &  6.178 $\pm$  0.001 &  5.251 $\pm$  0.009 \\
              2  &      054713.848+000016.92   &  F08/FA09  &     II?   & 11.748    & 11.129    & 10.129    &  9.310    &  8.728    &  8.233 $\pm$  0.003 &  7.824 $\pm$  0.003 &  7.599 $\pm$  0.001 &  7.022 $\pm$  0.001 &  4.531 $\pm$  0.014 \\
              3  &      054619.061+000329.56   &  F08/FA09  &    Flat   & 12.530    & 11.655    & 10.414    & 11.383 $\pm$  0.001 & 10.029 $\pm$  0.001 &  6.542 $\pm$  0.002 &  6.005 $\pm$  0.002 &  5.597 $\pm$  0.003 &  4.937 $\pm$  0.002 &  2.231 $\pm$  0.013 \\
              4  &      054811.880-002157.46   &            &    III?   & 11.994    & 11.216    & 10.427    &  9.418    &  9.016    &  8.981 $\pm$  0.005 &  8.972 $\pm$  0.002 &  8.841 $\pm$  0.003 &  8.758 $\pm$  0.004 &  8.593 $\pm$  0.145 \\
              5  &      054619.470-000519.97   &  F08/FA09  &     II?   & 11.986    &  --  & 10.602    &  9.668    &  9.127    &  7.960 $\pm$  0.003 &  7.497 $\pm$  0.002 &  7.151 $\pm$  0.002 &  6.052 $\pm$  0.003 &  3.002 $\pm$  0.010 \\
              6  &      054705.993+003208.41   &  F08/FA09  &     II?   & 13.361 $\pm$  0.001 & 12.570    & 10.705    & 10.013    &  9.275    &  8.095 $\pm$  0.003 &  7.578 $\pm$  0.003 &  7.249 $\pm$  0.001 &  6.527 $\pm$  0.001 &  3.101 $\pm$  0.009 \\
              7  &      054604.671+000458.01   &  F08/FA09  &     II?   & 11.511    & 11.854    & 10.926    &  9.954    &  9.301    &  8.296 $\pm$  0.003 &  7.887 $\pm$  0.004 &  7.472 $\pm$  0.001 &  6.692 $\pm$  0.001 &  3.796 $\pm$  0.015 \\
              8  &      054618.893-000538.11   &  F08/FA09  &     II?   & 12.560    & 11.911    & 11.129    & 10.291    &  9.875    &  8.945 $\pm$  0.005 &  8.654 $\pm$  0.001 &  8.183 $\pm$  0.002 &  7.568 $\pm$  0.002 &  5.730 $\pm$  0.013 \\
\end{longtable}
\end{landscape}

\end{appendix}

\end{document}